\documentclass[acmsmall,sigplan,]{acmart}

\acmConference[GPCE]{GPCE}{November, 2020}{Illinois, USA}
\acmYear{2020}
\acmISBN{} 
\acmDOI{} 
\startPage{1}

\setcopyright{none}

\usepackage{xurl}
\usepackage{booktabs}   
\usepackage{subcaption} 
\usepackage{tikz-cd}

\usepackage{listings}

\providecommand{\tightlist}{%
  \setlength{\itemsep}{0pt}\setlength{\parskip}{0pt}}

\usepackage{fancyvrb}

\DefineVerbatimEnvironment{Highlighting}{Verbatim}{commandchars=\\\{\}}

\newlength{\cslhangindent}
\newenvironment{cslreferences}%
  {\setlength{\parindent}{0pt}%
  \everypar{\setlength{\hangindent}{\cslhangindent}}\ignorespaces}%
  {\par}

\begin{document}

\title[Less Arbitrary]{Less Arbitrary
waiting time}
\providecommand{\subtitle}[1]{}
\subtitle{Short paper}
    \author{Michał J. Gajda}
      \date{2020-05-05}
\title{Less Arbitrary waiting
time}         
\subtitle{Short
paper}                     


\begin{abstract}
Property testing is the cheapest and
most precise way of building up a test
suite for your program. Especially if
the datatypes enjoy nice mathematical
laws. But it is also the easiest way to
make it run for an unreasonably long
time. We prove connection between deeply
recursive data structures, and epidemic
growth rate, and show how to fix the
problem, and make Arbitrary instances
run in linear time with respect to
assumed test size.
\end{abstract}



\maketitle

\hypertarget{introduction}{%
\section{Introduction}\label{introduction}}

Property testing is the cheapest and
most precise way of building up a test
suite for your program. Especially if
the datatypes enjoy nice mathematical
laws. But it is also the easiest way to
make it run for an unreasonably long
time. We show that connection between
deeply recursive data structures, and
epidemic growth rate can be easily fixed
with a generic implementation. After our
intervention the Arbitrary instances run
in linear time with respect to assumed
test size. We also provide a fully
generic implementation, so error-prone
coding process is removed.

\hypertarget{motivation}{%
\section{Motivation}\label{motivation}}

Typical arbitrary instance just draws a
random constructor from a set, possibly
biasing certain outcomes.

\textbf{Generic} arbitrary instance
looks like this:

\setlength{\tabcolsep}{1pt}
\begin{tabular}{llllllllllll}
\multicolumn{3}{l}{$\textbf{data}$} & \multicolumn{4}{l}{$\textit{ }\textsc{Tree}\textit{        }$} & \multicolumn{1}{l}{$\alpha$} & \multicolumn{2}{l}{$\textit{ }\textit{=}\textit{
}$}\\
\multicolumn{3}{p{4ex}}{    } & \multicolumn{4}{l}{$\textsc{Leaf}\textit{         }$} & \multicolumn{1}{l}{$\alpha$} & \multicolumn{2}{l}{$\textit{
}$}\\
\multicolumn{2}{p{2ex}}{  } & \multicolumn{1}{l}{$\textit{|}\textit{ }$} & \multicolumn{1}{l}{$\textsc{Branch}$} & \multicolumn{1}{l}{$\textit{ }$} & \multicolumn{1}{l}{$\textit{[}$} & \multicolumn{1}{l}{$\textsc{Tree}\textit{ }$} & \multicolumn{1}{l}{$\alpha$} & \multicolumn{1}{l}{$\textit{]}$} & \multicolumn{1}{l}{$\textit{
}$}\\
\multicolumn{2}{p{2ex}}{  } & \multicolumn{2}{l}{$\textbf{deriving}$} & \multicolumn{1}{l}{$\textit{ }$} & \multicolumn{1}{l}{$\textit{(}$} & \multicolumn{3}{l}{$\textsc{Eq}\textit{,}\textsc{Show}$} & \multicolumn{1}{l}{$\textit{,}\textsc{Generic.Generic}\textit{)}$}\\

\end{tabular}

\setlength{\tabcolsep}{1pt}
\begin{tabular}{llllllllllllllllll}
\multicolumn{4}{l}{$\textbf{instance}$} & \multicolumn{1}{l}{$\textit{ }$} & \multicolumn{1}{l}{$\textsc{Arbitrary}$} & \multicolumn{4}{l}{$\textit{       }$} & \multicolumn{1}{l}{$\alpha$} & \multicolumn{5}{l}{$\textit{
}$}\\
\multicolumn{3}{p{6ex}}{      } & \multicolumn{1}{l}{$\textit{=>}$} & \multicolumn{1}{l}{$\textit{ }$} & \multicolumn{1}{l}{$\textsc{Arbitrary}$} & \multicolumn{1}{l}{$\textit{ }$} & \multicolumn{1}{l}{$\textit{(}$} & \multicolumn{2}{l}{$\textsc{Tree}\textit{ }$} & \multicolumn{1}{l}{$\alpha$} & \multicolumn{1}{l}{$\textit{)}\textit{ }$} & \multicolumn{4}{l}{$\textbf{where}\textit{
}$}\\
\multicolumn{2}{p{2ex}}{  } & \multicolumn{5}{l}{$\emph{arbitrary}\textit{ }\textit{=}\textit{ }\emph{oneof}$} & \multicolumn{1}{l}{$\textit{ }$} & \multicolumn{1}{l}{$\textit{[}$} & \multicolumn{1}{l}{$\textsc{Leaf}$} & \multicolumn{2}{l}{$\textit{   }$} & \multicolumn{1}{l}{$\mathbin{\ooalign{\raise.29ex\hbox{$\scriptscriptstyle\$$}\cr\hss$\!\lozenge$\hss}}$} & \multicolumn{1}{l}{$\textit{ }$} & \multicolumn{1}{l}{$\emph{arbitrary}$} & \multicolumn{1}{l}{$\textit{
}$}\\
\multicolumn{8}{p{20ex}}{                    } & \multicolumn{1}{l}{$\textit{,}$} & \multicolumn{3}{l}{$\textsc{Branch}\textit{ }$} & \multicolumn{1}{l}{$\mathbin{\ooalign{\raise.29ex\hbox{$\scriptscriptstyle\$$}\cr\hss$\!\lozenge$\hss}}$} & \multicolumn{1}{l}{$\textit{ }$} & \multicolumn{1}{l}{$\emph{arbitrary}$} & \multicolumn{1}{l}{$\textit{
}$}\\
\multicolumn{8}{p{20ex}}{                    } & \multicolumn{8}{l}{$\textit{]}$}\\

\end{tabular}

Assuming we run QuickCheck with any size
parameter greater than 1, it will fail
to terminate!

List instance is a wee bit better, since
it tries to limit maximum list length to
a constant option:

\setlength{\tabcolsep}{1pt}
\begin{tabular}{llllllllllllllll}
\multicolumn{5}{l}{$\textbf{instance}$} & \multicolumn{1}{l}{$\textit{ }$} & \multicolumn{2}{l}{$\textsc{Arbitrary}$} & \multicolumn{1}{l}{$\textit{  }$} & \multicolumn{1}{l}{$\alpha$} & \multicolumn{4}{l}{$\textit{
}$}\\
\multicolumn{4}{p{6ex}}{      } & \multicolumn{1}{l}{$\textit{=>}$} & \multicolumn{1}{l}{$\textit{ }$} & \multicolumn{2}{l}{$\textsc{Arbitrary}$} & \multicolumn{1}{l}{$\textit{ }\textit{[}$} & \multicolumn{1}{l}{$\alpha$} & \multicolumn{1}{l}{$\textit{]}\textit{ }$} & \multicolumn{3}{l}{$\textbf{where}\textit{
}$}\\
\multicolumn{2}{p{2ex}}{  } & \multicolumn{6}{l}{$\emph{lessArbitrary}\textit{ }\textit{=}\textit{ }$} & \multicolumn{3}{l}{$\emph{sized}$} & \multicolumn{1}{l}{$\textit{ }\textit{\$}\textit{ }$} & \multicolumn{1}{l}{$\textit{\textbackslash{}}$} & \multicolumn{1}{l}{$\emph{size}\textit{ }\textbf{do}\textit{
}$}\\
\multicolumn{3}{p{4ex}}{    } & \multicolumn{4}{l}{$\emph{len}\textit{  }\textit{<-}\textit{ }$} & \multicolumn{1}{l}{$\emph{choose}$} & \multicolumn{4}{l}{$\textit{ }\textit{(}1\textit{,}\emph{size}$} & \multicolumn{1}{l}{$\textit{)}$} & \multicolumn{1}{l}{$\textit{
}$}\\
\multicolumn{3}{p{4ex}}{    } & \multicolumn{4}{l}{$\emph{vectorOf}$} & \multicolumn{7}{l}{$\textit{ }\emph{len}\textit{ }\emph{lessArbitrary}$}\\

\end{tabular}

Indeed QuickCheck
manual{[}\protect\hyperlink{ref-quickcheck-manual}{10}{]},
suggests an error-prone, manual method
of limiting the depth of generated
structure by dividing \texttt{size} by
reproduction factor of the
structure\footnote{We changed
  \texttt{liftM} and \texttt{liftM2}
  operators to
  \texttt{\textless{}\$\textgreater{}}
  and
  \texttt{\textless{}*\textgreater{}}
  for clarity and consistency.} :

\setlength{\tabcolsep}{1pt}
\begin{tabular}{llllllllllllllllllllllll}
\multicolumn{22}{l}{$\textbf{data}\textit{ }\textsc{Tree}\textit{ }\textit{=}\textit{ }\textsc{Leaf}\textit{ }\textsc{Int}\textit{ }\textit{|}\textit{ }\textsc{Branch}\textit{ }\textsc{Tree}\textit{ }\textsc{Tree}\textit{
}$}\\
\multicolumn{22}{l}{$\textit{}$}\\
\multicolumn{13}{l}{$\textbf{instance}\textit{ }\textsc{Arbitrary}\textit{ }$} & \multicolumn{9}{l}{$\textsc{Tree}\textit{ }\textbf{where}\textit{
}$}\\
\multicolumn{2}{p{2ex}}{  } & \multicolumn{11}{l}{$\emph{arbitrary}\textit{ }\textit{=}\textit{ }\emph{sized}$} & \multicolumn{1}{l}{$\textit{ }$} & \multicolumn{3}{l}{$\emph{tree'}$} & \multicolumn{5}{l}{$\textit{
}$}\\
\multicolumn{3}{p{4ex}}{    } & \multicolumn{6}{l}{$\textbf{where}\textit{ }\emph{tree'}$} & \multicolumn{1}{l}{$\textit{ }$} & \multicolumn{1}{l}{$0$} & \multicolumn{1}{l}{$\textit{ }$} & \multicolumn{1}{l}{$\textit{=}$} & \multicolumn{1}{l}{$\textit{ }$} & \multicolumn{3}{l}{$\textsc{Leaf}\textit{ }$} & \multicolumn{5}{l}{$\mathbin{\ooalign{\raise.29ex\hbox{$\scriptscriptstyle\$$}\cr\hss$\!\lozenge$\hss}}\textit{ }\emph{arbitrary}\textit{
}$}\\
\multicolumn{4}{p{6ex}}{      } & \multicolumn{3}{l}{$\emph{tree'}\textit{ }\emph{n}$} & \multicolumn{1}{l}{$\textit{ }$} & \multicolumn{1}{l}{$\textit{|}$} & \multicolumn{1}{l}{$\textit{ }$} & \multicolumn{1}{l}{$\emph{n}$} & \multicolumn{1}{l}{$\textit{>}$} & \multicolumn{1}{l}{$0$} & \multicolumn{1}{l}{$\textit{ }$} & \multicolumn{8}{l}{$\textit{=}\textit{ 
}$}\\
\multicolumn{5}{p{8ex}}{        } & \multicolumn{2}{l}{$\emph{oneof}$} & \multicolumn{1}{l}{$\textit{ }$} & \multicolumn{1}{l}{$\textit{[}$} & \multicolumn{4}{l}{$\textsc{Leaf}$} & \multicolumn{3}{l}{$\textit{   }$} & \multicolumn{6}{l}{$\mathbin{\ooalign{\raise.29ex\hbox{$\scriptscriptstyle\$$}\cr\hss$\!\lozenge$\hss}}\textit{ }\emph{arbitrary}\textit{,}\textit{
}$}\\
\multicolumn{6}{p{11ex}}{           } & \multicolumn{9}{l}{$\textsc{Branch}\textit{ }\mathbin{\ooalign{\raise.29ex\hbox{$\scriptscriptstyle\$$}\cr\hss$\!\lozenge$\hss}}$} & \multicolumn{1}{l}{$\textit{ }$} & \multicolumn{2}{l}{$\emph{subtree}$} & \multicolumn{1}{l}{$\textit{ }$} & \multicolumn{1}{l}{$\mathbin{\ooalign{\raise.37ex\hbox{$\scriptscriptstyle{*}$}\cr\hss$\!\lozenge$\hss}}$} & \multicolumn{1}{l}{$\textit{ }$} & \multicolumn{1}{l}{$\emph{subtree}\textit{]}\textit{
}$}\\
\multicolumn{5}{p{8ex}}{        } & \multicolumn{10}{l}{$\textbf{where}\textit{ }\emph{subtree}$} & \multicolumn{1}{l}{$\textit{ }$} & \multicolumn{2}{l}{$\textit{=}\textit{ }\emph{tree'}$} & \multicolumn{1}{l}{$\textit{ }$} & \multicolumn{1}{l}{$\textit{(}\emph{n}\textit{ }$} & \multicolumn{1}{l}{$\textit{`}$} & \multicolumn{1}{l}{$\emph{div}\textit{`}\textit{ }2\textit{)}$}\\

\end{tabular}

Above example uses division of size by
maximum branching factor to decrease
coverage into relatively deep data
structures, whereas dividing by average
branching factor of
\texttt{\textasciitilde{}2} will
generate both deep and very large
structures.

This fixes non-termination issue, but
still may lead to unpredictable waiting
times for nested structures. The depth
of the generated structure is linearly
limited by dividing the \texttt{n} by
expected branching factor of the
recursive data structure. However this
does not work very well for mutually
recursive data structures occuring in
compilers{[}\protect\hyperlink{ref-compilersALaCarte}{3}{]},
which may have 30 constructors with
highly variable\footnote{Due to list
  parameters.} branching factor just
like GHC's \texttt{HSExpr} data types.

Now we have a choice of manual
generation of these data structures,
which certainly introduces bias in
testing, or abandoning property testing
for real-life-sized projects.

\hypertarget{complexity-analysis}{%
\section{Complexity
analysis}\label{complexity-analysis}}

We might be tempted to compute average
size of the structure. Let's use
reproduction rate estimate for a single
rewrite of \texttt{arbitrary} function
written in conventional way.

We compute a number of recursive
references for each constructor. Then we
take an average number of references
among all the constructors. If it is
greater than 1, any \textbf{non-lazy}
property test will certainly fail to
terminate. If it is slightly smaller, we
still can wait a long time.

What is an issue here is not just
non-termination which is fixed by
error-prone manual process of writing
own instances that use explicit
\texttt{size} parameter.

The much worse issue is unpredictability
of the test runtime. Final issue is the
poor coverage for mutually recursive
data structure with multitude of
constructors.

Given a \emph{maximum size} parameter
(as it is now called) to QuickCheck,
would we not expect that tests terminate
within linear time of this parameter? At
least if our computation algorithms are
linear with respect to input size?

Currently for any recursive structure
like \texttt{Tree\ a}, we see some
exponential function. For example
\(size^n\), where \(n\) is a random
variable.

\hypertarget{solution}{%
\section{Solution}\label{solution}}

We propose to replace implementation
with a simple state
monad{[}\protect\hyperlink{ref-composing-monads}{7}{]}
that actually remembers how many
constructors were generated, and thus
avoid limiting the depth of generated
data structures, and ignoring estimation
of branching factor altogether.

\setlength{\tabcolsep}{1pt}
\begin{tabular}{llllllllll}
\multicolumn{3}{l}{$\textbf{newtype}\textit{ }\textsc{Cost}$} & \multicolumn{1}{l}{$\textit{ }\textit{=}$} & \multicolumn{1}{l}{$\textit{ }$} & \multicolumn{1}{l}{$\textsc{Cost}$} & \multicolumn{1}{l}{$\textit{ }\textsc{Int}$} & \multicolumn{1}{l}{$\textit{
}$}\\
\multicolumn{2}{p{2ex}}{  } & \multicolumn{1}{l}{$\textbf{deriving}\textit{ }\textit{(}$} & \multicolumn{1}{l}{$\textsc{Eq}$} & \multicolumn{1}{l}{$\textit{,}$} & \multicolumn{1}{l}{$\textsc{Ord}\textit{,}$} & \multicolumn{1}{l}{$\textsc{Enum}$} & \multicolumn{1}{l}{$\textit{,}\textsc{Bounded}\textit{,}\textsc{Num}\textit{)}$}\\

\end{tabular}

\setlength{\tabcolsep}{1pt}
\begin{tabular}{lllllllllll}
\multicolumn{3}{l}{$\textbf{newtype}\textit{ }$} & \multicolumn{2}{l}{$\textsc{CostGen}$} & \multicolumn{4}{l}{$\textit{                               }\alpha\textit{ }\textit{=}\textit{
}$}\\
\multicolumn{3}{p{8ex}}{        } & \multicolumn{2}{l}{$\textsc{CostGen}$} & \multicolumn{4}{l}{$\textit{ }\textit{\{}\textit{
}$}\\
\multicolumn{4}{p{10ex}}{          } & \multicolumn{2}{l}{$\emph{runCostGen}$} & \multicolumn{1}{l}{$\textit{ }$} & \multicolumn{1}{l}{$\textit{::}\textit{ }\textsc{State.StateT}\textit{ }\textsc{Cost}$} & \multicolumn{1}{l}{$\textit{ }\textsc{QC.Gen}\textit{ }\alpha\textit{ }\textit{\}}\textit{
}$}\\
\multicolumn{2}{p{2ex}}{  } & \multicolumn{2}{l}{$\textbf{deriving}$} & \multicolumn{2}{l}{$\textit{ }\textit{(}\textsc{Functor}\textit{,}$} & \multicolumn{1}{l}{$\textit{ }$} & \multicolumn{1}{l}{$\textsc{Applicative}\textit{,}\textit{ }\textsc{Monad}\textit{,}\textit{ }$} & \multicolumn{1}{l}{$\textsc{State.MonadFix}\textit{)}$}\\

\end{tabular}

We track the spending in the usual way:

\setlength{\tabcolsep}{1pt}
\begin{tabular}{lllllllllll}
\multicolumn{3}{l}{$\emph{spend}$} & \multicolumn{1}{l}{$\textit{ }$} & \multicolumn{1}{l}{$\textit{::}$} & \multicolumn{1}{l}{$\textit{ }$} & \multicolumn{3}{l}{$\textsc{Cost}\textit{ }\textit{->}\textit{ }\textsc{CostGen}\textit{ }\textit{(}\textit{)}\textit{}$}\\
\multicolumn{3}{l}{$\emph{spend}$} & \multicolumn{1}{l}{$\textit{ }$} & \multicolumn{1}{l}{$\gamma\textit{ }$} & \multicolumn{1}{l}{$\textit{=}$} & \multicolumn{1}{l}{$\textit{ }$} & \multicolumn{1}{l}{$\textbf{do}$} & \multicolumn{1}{l}{$\textit{
}$}\\
\multicolumn{2}{p{2ex}}{  } & \multicolumn{4}{l}{$\textsc{CostGen}$} & \multicolumn{1}{l}{$\textit{ }$} & \multicolumn{1}{l}{$\textit{\$}\textit{ }$} & \multicolumn{1}{l}{$\emph{State.modify}\textit{ }\textit{(}\textit{-}\gamma\textit{+}\textit{)}\textit{
}$}\\
\multicolumn{2}{p{2ex}}{  } & \multicolumn{7}{l}{$\emph{checkBudget}$}\\

\end{tabular}

To make generation easier, we introduce
\texttt{budget\ check} operator:

\setlength{\tabcolsep}{1pt}
\begin{tabular}{lllllllllllllllllllll}
\multicolumn{5}{l}{$\textit{(}\textit{\$\$\$?}\textit{)}\textit{ }$} & \multicolumn{1}{l}{$\textit{::}$} & \multicolumn{1}{l}{$\textit{ }$} & \multicolumn{12}{l}{$\textsc{HasCallStack}\textit{
}$}\\
\multicolumn{5}{p{7ex}}{       } & \multicolumn{1}{l}{$\textit{=>}$} & \multicolumn{1}{l}{$\textit{ }$} & \multicolumn{5}{l}{$\textsc{CostGen}$} & \multicolumn{1}{l}{$\textit{ }$} & \multicolumn{1}{l}{$\alpha$} & \multicolumn{5}{l}{$\textit{
}$}\\
\multicolumn{5}{p{7ex}}{       } & \multicolumn{1}{l}{$\textit{->}$} & \multicolumn{1}{l}{$\textit{ }$} & \multicolumn{5}{l}{$\textsc{CostGen}$} & \multicolumn{1}{l}{$\textit{ }$} & \multicolumn{1}{l}{$\alpha$} & \multicolumn{5}{l}{$\textit{
}$}\\
\multicolumn{5}{p{7ex}}{       } & \multicolumn{1}{l}{$\textit{->}$} & \multicolumn{1}{l}{$\textit{ }$} & \multicolumn{5}{l}{$\textsc{CostGen}$} & \multicolumn{1}{l}{$\textit{ }$} & \multicolumn{1}{l}{$\alpha$} & \multicolumn{5}{l}{$\textit{}$}\\
\multicolumn{13}{l}{$\emph{cheapVariants}\textit{ }\textit{\$\$\$?}$} & \multicolumn{1}{l}{$\textit{ }$} & \multicolumn{5}{l}{$\emph{costlyVariants}\textit{ }\textit{=}\textit{ }\textbf{do}\textit{
}$}\\
\multicolumn{2}{p{2ex}}{  } & \multicolumn{12}{l}{$\emph{budget}\textit{ }\textit{<-}\textit{ }\textsc{CostGen}$} & \multicolumn{5}{l}{$\textit{ }\emph{State.get}\textit{
}$}\\
\multicolumn{2}{p{2ex}}{  } & \multicolumn{1}{l}{$\textbf{if}\textit{ }$} & \multicolumn{1}{l}{$\textit{|}$} & \multicolumn{1}{l}{$\textit{ }$} & \multicolumn{3}{l}{$\emph{budget}$} & \multicolumn{1}{l}{$\textit{ }$} & \multicolumn{1}{c}{$\textit{>}$} & \multicolumn{1}{l}{$\textit{ }$} & \multicolumn{3}{l}{$\textit{(}0\textit{ }$} & \multicolumn{1}{l}{$\textit{::}\textit{ }$} & \multicolumn{1}{l}{$\textsc{Cost}\textit{)}\textit{ }$} & \multicolumn{1}{l}{$\textit{->}$} & \multicolumn{1}{l}{$\textit{ }$} & \multicolumn{1}{l}{$\emph{costlyVariants}\textit{
}$}\\
\multicolumn{3}{p{5ex}}{     } & \multicolumn{1}{l}{$\textit{|}$} & \multicolumn{1}{l}{$\textit{ }$} & \multicolumn{3}{l}{$\emph{budget}$} & \multicolumn{1}{l}{$\textit{ }$} & \multicolumn{1}{c}{$\textit{>}$} & \multicolumn{1}{l}{$\textit{ }$} & \multicolumn{4}{l}{$-10000$} & \multicolumn{1}{l}{$\textit{      }$} & \multicolumn{1}{l}{$\textit{->}$} & \multicolumn{1}{l}{$\textit{ }$} & \multicolumn{1}{l}{$\emph{cheapVariants}\textit{
}$}\\
\multicolumn{3}{p{5ex}}{     } & \multicolumn{1}{l}{$\textit{|}$} & \multicolumn{1}{l}{$\textit{ }$} & \multicolumn{6}{l}{$\emph{otherwise}$} & \multicolumn{5}{l}{$\textit{            }$} & \multicolumn{1}{l}{$\textit{->}$} & \multicolumn{1}{l}{$\textit{ }$} & \multicolumn{1}{l}{$\emph{error}\textit{ }\textit{\$}\textit{
}$}\\
\multicolumn{5}{p{7ex}}{       } & \multicolumn{14}{l}{$\textit{"Recursive structure with no loop breaker."}$}\\

\end{tabular}

\setlength{\tabcolsep}{1pt}
\begin{tabular}{lllllllllllll}
\multicolumn{6}{l}{$\emph{checkBudget}$} & \multicolumn{1}{l}{$\textit{ }$} & \multicolumn{1}{l}{$\textit{::}$} & \multicolumn{3}{l}{$\textit{ }\textsc{HasCallStack}\textit{ }\textit{=>}\textit{ }\textsc{CostGen}\textit{ }\textit{(}\textit{)}\textit{}$}\\
\multicolumn{6}{l}{$\emph{checkBudget}$} & \multicolumn{1}{l}{$\textit{ }$} & \multicolumn{1}{l}{$\textit{=}\textit{ }$} & \multicolumn{3}{l}{$\textbf{do}\textit{
}$}\\
\multicolumn{2}{p{2ex}}{  } & \multicolumn{4}{l}{$\emph{budget}\textit{ }\textit{<-}$} & \multicolumn{1}{l}{$\textit{ }$} & \multicolumn{3}{l}{$\textsc{CostGen}\textit{ }$} & \multicolumn{1}{l}{$\emph{State.get}\textit{
}$}\\
\multicolumn{2}{p{2ex}}{  } & \multicolumn{1}{l}{$\textbf{if}$} & \multicolumn{3}{l}{$\textit{ }\emph{budget}$} & \multicolumn{1}{l}{$\textit{ }$} & \multicolumn{1}{l}{$\textit{<}\textit{ }$} & \multicolumn{2}{l}{$-10000$} & \multicolumn{1}{l}{$\textit{
}$}\\
\multicolumn{3}{p{4ex}}{    } & \multicolumn{1}{l}{$\textbf{then}$} & \multicolumn{1}{l}{$\textit{ }$} & \multicolumn{3}{l}{$\emph{error}$} & \multicolumn{1}{l}{$\textit{ }$} & \multicolumn{2}{l}{$\textit{"Recursive structure with no loop breaker."}\textit{
}$}\\
\multicolumn{3}{p{4ex}}{    } & \multicolumn{1}{l}{$\textbf{else}$} & \multicolumn{1}{l}{$\textit{ }$} & \multicolumn{4}{l}{$\emph{return}$} & \multicolumn{2}{l}{$\textit{ }\textit{(}\textit{)}$}\\

\end{tabular}

In order to conveniently define our
budget generators, we might want to
define a class for them:

\setlength{\tabcolsep}{1pt}
\begin{tabular}{lllllll}
\multicolumn{3}{l}{$\textbf{class}\textit{ }\textsc{LessArbitrary}$} & \multicolumn{1}{l}{$\textit{ }\alpha\textit{ }\textbf{where}$} & \multicolumn{1}{l}{$\textit{
}$}\\
\multicolumn{2}{p{2ex}}{  } & \multicolumn{1}{l}{$\emph{lessArbitrary}\textit{ }\textit{::}\textit{ }$} & \multicolumn{1}{l}{$\textsc{CostGen}\textit{ }$} & \multicolumn{1}{l}{$\alpha$}\\

\end{tabular}

\setlength{\tabcolsep}{1pt}
\begin{tabular}{llllllllll}
\multicolumn{1}{p{2ex}}{  } & \multicolumn{2}{l}{$\textbf{default}\textit{ }\emph{lessArbitrary}\textit{ }\textit{::}\textit{ }$} & \multicolumn{1}{l}{$\textit{(}$} & \multicolumn{1}{l}{$\textsc{Generic}\textit{             }$} & \multicolumn{1}{l}{$\alpha$} & \multicolumn{1}{l}{$\textit{
}$}\\
\multicolumn{3}{p{27ex}}{                           } & \multicolumn{1}{l}{$\textit{,}$} & \multicolumn{1}{l}{$\textsc{GLessArbitrary}\textit{ }\textit{(}\textsc{Rep}\textit{ }$} & \multicolumn{1}{l}{$\alpha$} & \multicolumn{1}{l}{$\textit{)}\textit{)}\textit{
}$}\\
\multicolumn{2}{p{24ex}}{                        } & \multicolumn{2}{l}{$\textit{=>}\textit{  }$} & \multicolumn{1}{l}{$\textsc{CostGen}\textit{             }$} & \multicolumn{1}{l}{$\alpha$} & \multicolumn{1}{l}{$\textit{
}$}\\
\multicolumn{1}{p{2ex}}{  } & \multicolumn{6}{l}{$\emph{lessArbitrary}\textit{ }\textit{=}\textit{ }\emph{genericLessArbitrary}$}\\

\end{tabular}

Then we can use them as implementation
of \texttt{arbitrary} that should have
been always used:

\setlength{\tabcolsep}{1pt}
\begin{tabular}{llllllllllll}
\multicolumn{4}{l}{$\emph{fasterArbitrary}$} & \multicolumn{1}{l}{$\textit{ }$} & \multicolumn{1}{l}{$\textit{::}$} & \multicolumn{3}{l}{$\textit{ }\textsc{LessArbitrary}\textit{ }\alpha\textit{ }\textit{=>}\textit{ }\textsc{QC.Gen}\textit{ }\alpha\textit{}$}\\
\multicolumn{4}{l}{$\emph{fasterArbitrary}$} & \multicolumn{1}{l}{$\textit{ }$} & \multicolumn{1}{l}{$\textit{=}\textit{ }$} & \multicolumn{3}{l}{$\emph{sizedCost}\textit{ }\emph{lessArbitrary}\textit{
}$}\\
\multicolumn{9}{l}{$\textit{}$}\\
\multicolumn{1}{l}{$\emph{sizedCost}$} & \multicolumn{1}{l}{$\textit{ }$} & \multicolumn{1}{l}{$\textit{::}\textit{ }$} & \multicolumn{4}{l}{$\textsc{CostGen}\textit{ }\alpha\textit{ }\textit{->}$} & \multicolumn{1}{l}{$\textit{ }$} & \multicolumn{1}{l}{$\textsc{QC.Gen}\textit{ }\alpha\textit{}$}\\
\multicolumn{1}{l}{$\emph{sizedCost}$} & \multicolumn{1}{l}{$\textit{ }$} & \multicolumn{1}{l}{$\emph{gen}$} & \multicolumn{4}{l}{$\textit{ }\textit{=}\textit{ }\emph{QC.sized}\textit{ }$} & \multicolumn{1}{l}{$\textit{(}$} & \multicolumn{1}{l}{$\textit{`}\emph{withCost}\textit{`}\textit{ }\emph{gen}\textit{)}$}\\

\end{tabular}

Then we can implement \texttt{Arbitrary}
instances simply with:

\setlength{\tabcolsep}{1pt}
\begin{tabular}{llllllll}
\multicolumn{4}{l}{$\textbf{instance}$} & \multicolumn{1}{l}{$\textit{ }$} & \multicolumn{1}{l}{$\textit{\_{}}\textit{
}$}\\
\multicolumn{3}{p{6ex}}{      } & \multicolumn{1}{l}{$\textit{=>}$} & \multicolumn{1}{l}{$\textit{ }$} & \multicolumn{1}{l}{$\textsc{Arbitrary}\textit{ }\alpha\textit{ }\textbf{where}\textit{
}$}\\
\multicolumn{2}{p{2ex}}{  } & \multicolumn{4}{l}{$\emph{arbitrary}\textit{ }\textit{=}\textit{ }\emph{fasterArbitrary}$}\\

\end{tabular}

Of course we still need to define
\texttt{LessArbitrary}, but after seeing
how simple was a \texttt{Generic}
defintion \texttt{Arbitrary} we have a
hope that our implementation will be:

\setlength{\tabcolsep}{1pt}
\begin{tabular}{llll}
\multicolumn{1}{l}{$\textbf{instance}\textit{ }\textsc{LessArbitrary}\textit{ }\textbf{where}$}\\

\end{tabular}

That is - we hope that the the generic
implementation will take over.

\hypertarget{introduction-to-ghc-generics}{%
\section{Introduction to GHC
generics}\label{introduction-to-ghc-generics}}

Generics allow us to provide default
instance, by encoding any datatype into
its generic \texttt{Rep}resentation:

\setlength{\tabcolsep}{1pt}
\begin{tabular}{lllllllllllllllllll}
\multicolumn{4}{l}{$\textbf{instance}\textit{ }$} & \multicolumn{4}{l}{$\textsc{Generics}$} & \multicolumn{1}{l}{$\textit{ }\textit{(}$} & \multicolumn{2}{l}{$\textsc{Tree}$} & \multicolumn{1}{l}{$\textit{ }$} & \multicolumn{1}{l}{$\alpha$} & \multicolumn{3}{l}{$\textit{)}\textit{ }\textbf{where}$} & \multicolumn{1}{l}{$\textit{
}$}\\
\multicolumn{2}{p{2ex}}{  } & \multicolumn{1}{l}{$\emph{to}\textit{   }$} & \multicolumn{1}{l}{$\textit{::}$} & \multicolumn{1}{l}{$\textit{ }$} & \multicolumn{1}{l}{$\textsc{Tree}$} & \multicolumn{1}{l}{$\textit{ }$} & \multicolumn{1}{l}{$\alpha\textit{ }$} & \multicolumn{1}{l}{$\textit{->}$} & \multicolumn{1}{l}{$\textit{ }$} & \multicolumn{1}{l}{$\textsc{Rep}$} & \multicolumn{1}{l}{$\textit{ }$} & \multicolumn{1}{l}{$\textit{(}$} & \multicolumn{1}{l}{$\textsc{Tree}\textit{ }$} & \multicolumn{1}{l}{$\alpha$} & \multicolumn{1}{l}{$\textit{)}$} & \multicolumn{1}{l}{$\textit{
}$}\\
\multicolumn{2}{p{2ex}}{  } & \multicolumn{1}{l}{$\emph{from}\textit{ }$} & \multicolumn{1}{l}{$\textit{::}$} & \multicolumn{1}{l}{$\textit{ }$} & \multicolumn{1}{l}{$\textsc{Rep}\textit{ }$} & \multicolumn{1}{l}{$\textit{(}$} & \multicolumn{2}{l}{$\textsc{Tree}$} & \multicolumn{1}{l}{$\textit{ }$} & \multicolumn{1}{l}{$\alpha\textit{)}\textit{ }$} & \multicolumn{2}{l}{$\textit{->}$} & \multicolumn{1}{l}{$\textit{ }\textsc{Tree}$} & \multicolumn{1}{l}{$\textit{ }$} & \multicolumn{2}{l}{$\alpha$}\\

\end{tabular}

The secret to making a generic function
is to create a set of \texttt{instance}
declarations for each type family
constructor.

So let's examine \texttt{Rep}resentation
of our working example, and see how to
declare instances:

\begin{enumerate}
\def\labelenumi{\arabic{enumi}.}
\tightlist
\item
  First we see datatype metadata
  \texttt{D1} that shows where our type
  was defined:
\end{enumerate}

\setlength{\tabcolsep}{1pt}
\begin{tabular}{lllllllll}
\multicolumn{4}{l}{$\textbf{type}$} & \multicolumn{3}{l}{$\textit{ }\textbf{instance}\textit{ }\textsc{Rep}\textit{ }\textit{(}\textsc{Tree}\textit{ }\alpha\textit{)}\textit{ }\textit{=}\textit{
}$}\\
\multicolumn{2}{p{2ex}}{  } & \multicolumn{2}{l}{$\textsc{D1}$} & \multicolumn{3}{l}{$\textit{
}$}\\
\multicolumn{3}{p{3ex}}{   } & \multicolumn{1}{l}{$\textit{(}$} & \multicolumn{1}{l}{$\textit{'}\textsc{MetaData}\textit{ }$} & \multicolumn{2}{l}{$\textit{"Tree"}\textit{
}$}\\
\multicolumn{5}{p{14ex}}{              } & \multicolumn{1}{l}{$\textit{"Test.Arbitrary"}$} & \multicolumn{1}{l}{$\textit{
}$}\\
\multicolumn{5}{p{14ex}}{              } & \multicolumn{1}{l}{$\textit{"less-arbitrary"}$} & \multicolumn{1}{l}{$\textit{ }\textit{'}\textsc{False}\textit{)}$}\\

\end{tabular}

\begin{enumerate}
\def\labelenumi{\arabic{enumi}.}
\setcounter{enumi}{1}
\tightlist
\item
  Then we have constructor metadata
  \texttt{C1}:
\end{enumerate}

\setlength{\tabcolsep}{1pt}
\begin{tabular}{llllll}
\multicolumn{1}{p{7ex}}{       } & \multicolumn{1}{l}{$\textit{(}\textsc{C1}$} & \multicolumn{1}{l}{$\textit{
}$}\\
\multicolumn{2}{p{10ex}}{          } & \multicolumn{1}{l}{$\textit{(}\textit{'}\textsc{MetaCons}\textit{ }\textit{"Leaf"}\textit{ }\textit{'}\textsc{PrefixI}\textit{ }\textit{'}\textsc{False}\textit{)}$}\\

\end{tabular}

\begin{enumerate}
\def\labelenumi{\arabic{enumi}.}
\setcounter{enumi}{2}
\tightlist
\item
  Then we have metadata for each field
  selector within a constructor:
\end{enumerate}

\setlength{\tabcolsep}{1pt}
\begin{tabular}{llllllll}
\multicolumn{1}{p{10ex}}{          } & \multicolumn{1}{l}{$\textit{(}\textsc{S1}$} & \multicolumn{3}{l}{$\textit{
}$}\\
\multicolumn{2}{p{13ex}}{             } & \multicolumn{3}{l}{$\textit{(}\textit{'}\textsc{MetaSel}\textit{
}$}\\
\multicolumn{3}{p{16ex}}{                } & \multicolumn{1}{l}{$\textit{'}$} & \multicolumn{1}{l}{$\textsc{Nothing}\textit{
}$}\\
\multicolumn{3}{p{16ex}}{                } & \multicolumn{1}{l}{$\textit{'}$} & \multicolumn{1}{l}{$\textsc{NoSourceUnpackedness}\textit{
}$}\\
\multicolumn{3}{p{16ex}}{                } & \multicolumn{1}{l}{$\textit{'}$} & \multicolumn{1}{l}{$\textsc{NoSourceStrictness}\textit{
}$}\\
\multicolumn{3}{p{16ex}}{                } & \multicolumn{1}{l}{$\textit{'}$} & \multicolumn{1}{l}{$\textsc{DecidedLazy}\textit{)}$}\\

\end{tabular}

\begin{enumerate}
\def\labelenumi{\arabic{enumi}.}
\setcounter{enumi}{3}
\tightlist
\item
  And reference to another datatype in
  the record field value:
\end{enumerate}

\setlength{\tabcolsep}{1pt}
\begin{tabular}{lllll}
\multicolumn{1}{p{13ex}}{             } & \multicolumn{1}{l}{$\textit{(}\textsc{Rec0}\textit{ }\alpha\textit{)}\textit{)}$}\\

\end{tabular}

\begin{enumerate}
\def\labelenumi{\arabic{enumi}.}
\setcounter{enumi}{4}
\tightlist
\item
  Different constructors are joined by
  sum type operator:
\end{enumerate}

\setlength{\tabcolsep}{1pt}
\begin{tabular}{lllll}
\multicolumn{1}{p{8ex}}{        } & \multicolumn{1}{l}{$\textit{:+:}$}\\

\end{tabular}

\begin{enumerate}
\def\labelenumi{\arabic{enumi}.}
\setcounter{enumi}{5}
\tightlist
\item
  Second constructor has a similar
  representation:
\end{enumerate}

\setlength{\tabcolsep}{1pt}
\begin{tabular}{llllllllllll}
\multicolumn{1}{p{8ex}}{        } & \multicolumn{8}{l}{$\textsc{C1}\textit{
}$}\\
\multicolumn{2}{p{14ex}}{              } & \multicolumn{1}{l}{$\textit{(}$} & \multicolumn{6}{l}{$\textit{'}\textsc{MetaCons}\textit{ }\textit{"Branch"}\textit{ }\textit{'}\textsc{PrefixI}\textit{ }\textit{'}\textsc{False}\textit{)}\textit{
}$}\\
\multicolumn{2}{p{14ex}}{              } & \multicolumn{1}{l}{$\textit{(}$} & \multicolumn{1}{l}{$\textsc{S1}$} & \multicolumn{5}{l}{$\textit{
}$}\\
\multicolumn{4}{p{17ex}}{                 } & \multicolumn{5}{l}{$\textit{(}\textit{'}\textsc{MetaSel}\textit{
}$}\\
\multicolumn{5}{p{20ex}}{                    } & \multicolumn{1}{l}{$\textit{'}$} & \multicolumn{3}{l}{$\textsc{Nothing}\textit{
}$}\\
\multicolumn{5}{p{20ex}}{                    } & \multicolumn{1}{l}{$\textit{'}$} & \multicolumn{3}{l}{$\textsc{NoSourceUnpackedness}\textit{
}$}\\
\multicolumn{5}{p{20ex}}{                    } & \multicolumn{1}{l}{$\textit{'}$} & \multicolumn{3}{l}{$\textsc{NoSourceStrictness}\textit{
}$}\\
\multicolumn{5}{p{20ex}}{                    } & \multicolumn{1}{l}{$\textit{'}$} & \multicolumn{1}{l}{$\textsc{DecidedLazy}$} & \multicolumn{1}{l}{$\textit{)}$} & \multicolumn{1}{l}{$\textit{
}$}\\
\multicolumn{5}{p{20ex}}{                    } & \multicolumn{1}{l}{$\textit{(}$} & \multicolumn{1}{l}{$\textsc{Rec0}\textit{ }\textit{[}\textsc{Tree}\textit{ }$} & \multicolumn{1}{l}{$\alpha$} & \multicolumn{1}{l}{$\textit{]}\textit{)}\textit{)}\textit{)}\textit{
}$}\\
\multicolumn{6}{p{21ex}}{                     } & \multicolumn{3}{l}{$\emph{ignored}$}\\

\end{tabular}

\begin{enumerate}
\def\labelenumi{\arabic{enumi}.}
\setcounter{enumi}{6}
\tightlist
\item
  Note that \texttt{Rep}resentation type
  constructors have additional parameter
  that is not relevant for our use case.
\end{enumerate}

For simple datatypes, we are only
interested in three constructors:

\begin{itemize}
\tightlist
\item
  \texttt{:+:} encode choice between
  constructors
\item
  \texttt{:*:} encode a sequence of
  constructor parameters
\item
  \texttt{M1} encode metainformation
  about the named constructors,
  \texttt{C1}, \texttt{S1} and
  \texttt{D1} are actually shorthands
  for \texttt{M1\ C}, \texttt{M1\ S} and
  \texttt{M1\ D}
\end{itemize}

There are more short cuts to consider: *
\texttt{U1} is the unit type (no fields)
* \texttt{Rec0} is another type in the
field

\hypertarget{example-of-generics}{%
\subsection{Example of
generics}\label{example-of-generics}}

This generic representation can then be
matched by generic instances. Example of
\texttt{Arbitrary} instance from
{[}\protect\hyperlink{ref-generic-arbitrary}{5}{]}
serves as a basic example\footnote{We
  modified class name to simplify.}

\begin{enumerate}
\def\labelenumi{\arabic{enumi}.}
\tightlist
\item
  First we convert the type to its
  generic representation:
\end{enumerate}

\setlength{\tabcolsep}{1pt}
\begin{tabular}{llllllllll}
\multicolumn{4}{l}{$\emph{genericArbitrary}\textit{ }\textit{::}\textit{ }$} & \multicolumn{1}{l}{$\textit{(}$} & \multicolumn{1}{l}{$\textsc{Generic}\textit{        }$} & \multicolumn{1}{l}{$\alpha$} & \multicolumn{1}{l}{$\textit{
}$}\\
\multicolumn{4}{p{20ex}}{                    } & \multicolumn{1}{l}{$\textit{,}$} & \multicolumn{1}{l}{$\textsc{Arbitrary}\textit{ }\textit{(}\textsc{Rep}\textit{ }$} & \multicolumn{1}{l}{$\alpha$} & \multicolumn{1}{l}{$\textit{)}\textit{)}\textit{
}$}\\
\multicolumn{2}{p{17ex}}{                 } & \multicolumn{1}{l}{$\textit{=>}$} & \multicolumn{2}{l}{$\textit{  }$} & \multicolumn{1}{l}{$\textsc{Gen}\textit{            }$} & \multicolumn{1}{l}{$\alpha$} & \multicolumn{1}{l}{$\textit{}$}\\
\multicolumn{3}{l}{$\emph{genericArbitrary}\textit{  }\textit{=}$} & \multicolumn{5}{l}{$\textit{ }\emph{to}\textit{ }\mathbin{\ooalign{\raise.29ex\hbox{$\scriptscriptstyle\$$}\cr\hss$\!\lozenge$\hss}}\textit{ }\emph{arbitrary}$}\\

\end{tabular}

\begin{enumerate}
\def\labelenumi{\arabic{enumi}.}
\setcounter{enumi}{1}
\tightlist
\item
  We take care of nullary constructors
  with:
\end{enumerate}

\setlength{\tabcolsep}{1pt}
\begin{tabular}{lllllll}
\multicolumn{3}{l}{$\textbf{instance}\textit{ }\textsc{Arbitrary}$} & \multicolumn{1}{l}{$\textit{ }$} & \multicolumn{1}{l}{$\textsc{G.U1}\textit{ }\textbf{where}\textit{
}$}\\
\multicolumn{2}{p{2ex}}{  } & \multicolumn{1}{l}{$\emph{arbitrary}\textit{ }\textit{=}\textit{ }\emph{pure}$} & \multicolumn{1}{l}{$\textit{ }$} & \multicolumn{1}{l}{$\textsc{G.U1}$}\\

\end{tabular}

\begin{enumerate}
\def\labelenumi{\arabic{enumi}.}
\setcounter{enumi}{2}
\tightlist
\item
  For all fields arguments are
  recursively calling \texttt{Arbitrary}
  class method:
\end{enumerate}

\setlength{\tabcolsep}{1pt}
\begin{tabular}{lllllllll}
\multicolumn{3}{l}{$\textbf{instance}\textit{ }\textsc{Arbitrary}\textit{ }$} & \multicolumn{1}{l}{$\gamma$} & \multicolumn{1}{l}{$\textit{ }\textit{=>}$} & \multicolumn{1}{l}{$\textit{ }$} & \multicolumn{1}{l}{$\textsc{Arbitrary}\textit{ }\textit{(}\textsc{G.K1}\textit{ }\emph{i}\textit{ }\gamma\textit{)}\textit{ }\textbf{where}\textit{
}$}\\
\multicolumn{2}{p{2ex}}{  } & \multicolumn{1}{l}{$\emph{gArbitrary}\textit{ }\textit{=}\textit{ }\textsc{G.K1}$} & \multicolumn{1}{l}{$\textit{ }$} & \multicolumn{1}{l}{$\mathbin{\ooalign{\raise.29ex\hbox{$\scriptscriptstyle\$$}\cr\hss$\!\lozenge$\hss}}$} & \multicolumn{1}{l}{$\textit{ }$} & \multicolumn{1}{l}{$\emph{arbitrary}$}\\

\end{tabular}

\begin{enumerate}
\def\labelenumi{\arabic{enumi}.}
\setcounter{enumi}{3}
\tightlist
\item
  We skip metadata by the same recursive
  call:
\end{enumerate}

\setlength{\tabcolsep}{1pt}
\begin{tabular}{llllllllllll}
\multicolumn{4}{l}{$\textbf{instance}$} & \multicolumn{1}{l}{$\textit{ }$} & \multicolumn{1}{l}{$\textsc{Arbitrary}$} & \multicolumn{2}{l}{$\textit{           }$} & \multicolumn{1}{l}{$\emph{f}$} & \multicolumn{1}{l}{$\textit{
}$}\\
\multicolumn{3}{p{6ex}}{      } & \multicolumn{1}{l}{$\textit{=>}$} & \multicolumn{1}{l}{$\textit{ }$} & \multicolumn{1}{l}{$\textsc{Arbitrary}$} & \multicolumn{1}{l}{$\textit{ }$} & \multicolumn{1}{l}{$\textit{(}\textsc{G.M1}\textit{ }\emph{i}\textit{ }\gamma\textit{ }$} & \multicolumn{1}{l}{$\emph{f}$} & \multicolumn{1}{l}{$\textit{)}\textit{ }\textbf{where}\textit{
}$}\\
\multicolumn{2}{p{2ex}}{  } & \multicolumn{4}{l}{$\emph{arbitrary}\textit{ }\textit{=}\textit{ }\textsc{G.M1}$} & \multicolumn{1}{l}{$\textit{ }$} & \multicolumn{3}{l}{$\mathbin{\ooalign{\raise.29ex\hbox{$\scriptscriptstyle\$$}\cr\hss$\!\lozenge$\hss}}\textit{ }\emph{arbitrary}$}\\

\end{tabular}

\begin{enumerate}
\def\labelenumi{\arabic{enumi}.}
\setcounter{enumi}{4}
\tightlist
\item
  Given that all arguments of each
  constructor are joined by
  \texttt{:*:}, we need to recursively
  delve there too:
\end{enumerate}

\setlength{\tabcolsep}{1pt}
\begin{tabular}{lllllllllllllll}
\multicolumn{4}{l}{$\textbf{instance}\textit{ }$} & \multicolumn{1}{l}{$\textit{(}$} & \multicolumn{1}{l}{$\textsc{Arbitrary}$} & \multicolumn{7}{l}{$\textit{  }\alpha\textit{,}\textit{
}$}\\
\multicolumn{4}{p{9ex}}{         } & \multicolumn{1}{l}{$\textit{,}$} & \multicolumn{1}{l}{$\textsc{Arbitrary}$} & \multicolumn{4}{l}{$\textit{          }$} & \multicolumn{1}{l}{$\beta$} & \multicolumn{1}{l}{$\textit{)}$} & \multicolumn{1}{l}{$\textit{
}$}\\
\multicolumn{3}{p{6ex}}{      } & \multicolumn{2}{l}{$\textit{=>}\textit{  }$} & \multicolumn{1}{l}{$\textsc{Arbitrary}$} & \multicolumn{1}{l}{$\textit{ }$} & \multicolumn{1}{l}{$\textit{(}$} & \multicolumn{1}{l}{$\alpha$} & \multicolumn{1}{l}{$\textit{ }\textit{G.:*:}\textit{ }$} & \multicolumn{1}{l}{$\beta$} & \multicolumn{1}{l}{$\textit{)}$} & \multicolumn{1}{l}{$\textit{ }\textbf{where}\textit{
}$}\\
\multicolumn{2}{p{2ex}}{  } & \multicolumn{5}{l}{$\emph{arbitrary}\textit{ }\textit{=}\textit{ }\textit{(}\textit{G.:*:}$} & \multicolumn{1}{l}{$\textit{)}$} & \multicolumn{1}{l}{$\textit{ }$} & \multicolumn{4}{l}{$\mathbin{\ooalign{\raise.29ex\hbox{$\scriptscriptstyle\$$}\cr\hss$\!\lozenge$\hss}}\textit{ }\emph{arbitrary}\textit{ }\mathbin{\ooalign{\raise.37ex\hbox{$\scriptscriptstyle{*}$}\cr\hss$\!\lozenge$\hss}}\textit{ }\emph{arbitrary}$}\\

\end{tabular}

\begin{enumerate}
\def\labelenumi{\arabic{enumi}.}
\setcounter{enumi}{5}
\tightlist
\item
  In order to sample all constructors
  with the same probability we compute a
  number of constructor in each
  representation type with
  \texttt{SumLen} type family:
\end{enumerate}

\setlength{\tabcolsep}{1pt}
\begin{tabular}{llllllllllllllllll}
\multicolumn{6}{l}{$\textbf{type}\textit{ }\textbf{family}$} & \multicolumn{1}{l}{$\textit{ }$} & \multicolumn{1}{l}{$\textsc{SumLen}$} & \multicolumn{1}{l}{$\textit{ }$} & \multicolumn{1}{l}{$\alpha$} & \multicolumn{1}{l}{$\textit{ }$} & \multicolumn{2}{l}{$\textit{::}$} & \multicolumn{1}{l}{$\textit{ }$} & \multicolumn{1}{l}{$\textsc{Nat}\textit{ }\textbf{where}$} & \multicolumn{1}{l}{$\textit{
}$}\\
\multicolumn{2}{p{2ex}}{  } & \multicolumn{1}{l}{$\textsc{SumLen}$} & \multicolumn{1}{l}{$\textit{ }$} & \multicolumn{1}{l}{$\textit{(}$} & \multicolumn{1}{l}{$\alpha$} & \multicolumn{1}{l}{$\textit{ }$} & \multicolumn{1}{l}{$\textit{G.:+:}\textit{ }$} & \multicolumn{1}{l}{$\beta$} & \multicolumn{1}{l}{$\textit{)}$} & \multicolumn{1}{l}{$\textit{ }$} & \multicolumn{1}{l}{$\textit{=}$} & \multicolumn{1}{l}{$\textit{ }$} & \multicolumn{1}{l}{$\textit{(}$} & \multicolumn{1}{l}{$\textsc{SumLen}\textit{ }\alpha\textit{)}$} & \multicolumn{1}{l}{$\textit{ }\textit{+}\textit{ }\textit{(}\textsc{SumLen}\textit{ }\beta\textit{)}\textit{
}$}\\
\multicolumn{2}{p{2ex}}{  } & \multicolumn{1}{l}{$\textsc{SumLen}$} & \multicolumn{1}{l}{$\textit{ }$} & \multicolumn{1}{l}{$\alpha$} & \multicolumn{6}{l}{$\textit{           }$} & \multicolumn{1}{l}{$\textit{=}$} & \multicolumn{1}{l}{$\textit{ }$} & \multicolumn{3}{l}{$1$}\\

\end{tabular}

Now that we have number of constructors
computed, we can draw them with equal
probability:

\setlength{\tabcolsep}{1pt}
\begin{tabular}{llllllllllllllllllllll}
\multicolumn{7}{l}{$\textbf{instance}\textit{ }$} & \multicolumn{1}{l}{$\textit{(}$} & \multicolumn{5}{l}{$\textsc{Arbitrary}$} & \multicolumn{8}{l}{$\textit{        }\alpha\textit{
}$}\\
\multicolumn{7}{p{9ex}}{         } & \multicolumn{1}{l}{$\textit{,}$} & \multicolumn{5}{l}{$\textsc{Arbitrary}$} & \multicolumn{8}{l}{$\textit{                }\beta\textit{
}$}\\
\multicolumn{7}{p{9ex}}{         } & \multicolumn{1}{l}{$\textit{,}$} & \multicolumn{4}{l}{$\textsc{KnownNat}$} & \multicolumn{1}{l}{$\textit{ }$} & \multicolumn{2}{l}{$\textit{(}$} & \multicolumn{3}{l}{$\textsc{SumLen}$} & \multicolumn{3}{l}{$\textit{ }\alpha\textit{)}\textit{
}$}\\
\multicolumn{7}{p{9ex}}{         } & \multicolumn{1}{l}{$\textit{,}$} & \multicolumn{4}{l}{$\textsc{KnownNat}$} & \multicolumn{1}{l}{$\textit{ }$} & \multicolumn{2}{l}{$\textit{(}$} & \multicolumn{3}{l}{$\textsc{SumLen}$} & \multicolumn{3}{l}{$\textit{         }\beta\textit{)}\textit{
}$}\\
\multicolumn{7}{p{9ex}}{         } & \multicolumn{1}{l}{$\textit{)}$} & \multicolumn{13}{l}{$\textit{
}$}\\
\multicolumn{5}{p{6ex}}{      } & \multicolumn{3}{l}{$\textit{=>}\textit{  }$} & \multicolumn{13}{l}{$\textsc{Arbitrary}\textit{       }\textit{(}\alpha\textit{ }\textit{G.:+:}\textit{ }\beta\textit{)}\textit{ }\textbf{where}\textit{
}$}\\
\multicolumn{2}{p{2ex}}{  } & \multicolumn{7}{l}{$\emph{arbitrary}\textit{ }$} & \multicolumn{1}{l}{$\textit{=}$} & \multicolumn{1}{l}{$\textit{ }$} & \multicolumn{6}{l}{$\emph{frequency}$} & \multicolumn{4}{l}{$\textit{
}$}\\
\multicolumn{3}{p{4ex}}{    } & \multicolumn{1}{l}{$\textit{[}$} & \multicolumn{1}{l}{$\textit{ }$} & \multicolumn{1}{l}{$\textit{(}$} & \multicolumn{3}{l}{$\emph{lfreq}$} & \multicolumn{1}{l}{$\textit{,}$} & \multicolumn{1}{l}{$\textit{ }$} & \multicolumn{1}{l}{$\textsc{G.L1}$} & \multicolumn{1}{l}{$\textit{ }$} & \multicolumn{3}{c}{$\mathbin{\ooalign{\raise.29ex\hbox{$\scriptscriptstyle\$$}\cr\hss$\!\lozenge$\hss}}$} & \multicolumn{1}{l}{$\textit{ }$} & \multicolumn{2}{l}{$\emph{arbitrary}$} & \multicolumn{1}{l}{$\textit{)}$} & \multicolumn{1}{l}{$\textit{
}$}\\
\multicolumn{3}{p{4ex}}{    } & \multicolumn{1}{l}{$\textit{,}$} & \multicolumn{1}{l}{$\textit{ }$} & \multicolumn{1}{l}{$\textit{(}$} & \multicolumn{3}{l}{$\emph{rfreq}$} & \multicolumn{1}{l}{$\textit{,}$} & \multicolumn{1}{l}{$\textit{ }$} & \multicolumn{1}{l}{$\textsc{G.R1}$} & \multicolumn{1}{l}{$\textit{ }$} & \multicolumn{3}{c}{$\mathbin{\ooalign{\raise.29ex\hbox{$\scriptscriptstyle\$$}\cr\hss$\!\lozenge$\hss}}$} & \multicolumn{1}{l}{$\textit{ }$} & \multicolumn{2}{l}{$\emph{arbitrary}$} & \multicolumn{1}{l}{$\textit{)}$} & \multicolumn{1}{l}{$\textit{ }\textit{]}\textit{
}$}\\
\multicolumn{3}{p{4ex}}{    } & \multicolumn{18}{l}{$\textbf{where}\textit{
}$}\\
\multicolumn{5}{p{6ex}}{      } & \multicolumn{4}{l}{$\emph{lfreq}\textit{ }$} & \multicolumn{1}{l}{$\textit{=}$} & \multicolumn{1}{l}{$\textit{ }$} & \multicolumn{10}{l}{$\emph{fromIntegral}\textit{
}$}\\
\multicolumn{9}{p{12ex}}{            } & \multicolumn{1}{l}{$\textit{\$}$} & \multicolumn{1}{l}{$\textit{ }$} & \multicolumn{10}{l}{$\emph{natVal}\textit{ }\textit{(}\textsc{Proxy}\textit{ }\textit{::}\textit{ }\textsc{Proxy}\textit{ }\textit{(}\textsc{SumLen}\textit{ }\alpha\textit{)}\textit{)}\textit{
}$}\\
\multicolumn{5}{p{6ex}}{      } & \multicolumn{4}{l}{$\emph{rfreq}\textit{ }$} & \multicolumn{1}{l}{$\textit{=}$} & \multicolumn{1}{l}{$\textit{ }$} & \multicolumn{10}{l}{$\emph{fromIntegral}\textit{
}$}\\
\multicolumn{9}{p{12ex}}{            } & \multicolumn{1}{l}{$\textit{\$}$} & \multicolumn{1}{l}{$\textit{ }$} & \multicolumn{10}{l}{$\emph{natVal}\textit{ }\textit{(}\textsc{Proxy}\textit{ }\textit{::}\textit{ }\textsc{Proxy}\textit{ }\textit{(}\textsc{SumLen}\textit{ }\beta\textit{)}\textit{)}$}\\

\end{tabular}

Excellent piece of work, but
non-terminating for recursive types with
average branching factor greater than 1
(and non-lazy tests, like checking
\texttt{Eq} reflexivity.)

\hypertarget{implementing-with-generics}{%
\subsection{Implementing with
Generics}\label{implementing-with-generics}}

It is apparent from our previous
considerations, that we can reuse code
from the existing generic implementation
when the budget is positive. We just
need to spend a dollar for each
constructor we encounter.

For the \texttt{Monoid} the
implementation would be trivial, since
we can always use \texttt{mempty} and
assume it is cheap:

\setlength{\tabcolsep}{1pt}
\begin{tabular}{llllllllllll}
\multicolumn{5}{l}{$\emph{genericLessArbitraryMonoid}\textit{ }\textit{::}\textit{ }$} & \multicolumn{1}{l}{$\textit{(}$} & \multicolumn{1}{l}{$\textsc{Generic}\textit{             }$} & \multicolumn{1}{l}{$\alpha$} & \multicolumn{2}{l}{$\textit{
}$}\\
\multicolumn{5}{p{30ex}}{                              } & \multicolumn{1}{l}{$\textit{,}$} & \multicolumn{1}{l}{$\textsc{GLessArbitrary}\textit{ }\textit{(}\textsc{Rep}\textit{ }$} & \multicolumn{1}{l}{$\alpha$} & \multicolumn{1}{l}{$\textit{)}$} & \multicolumn{1}{l}{$\textit{
}$}\\
\multicolumn{5}{p{30ex}}{                              } & \multicolumn{1}{l}{$\textit{,}$} & \multicolumn{1}{l}{$\textsc{Monoid}\textit{              }$} & \multicolumn{1}{l}{$\alpha$} & \multicolumn{1}{l}{$\textit{ }$} & \multicolumn{1}{l}{$\textit{)}\textit{
}$}\\
\multicolumn{3}{p{27ex}}{                           } & \multicolumn{1}{l}{$\textit{=>}$} & \multicolumn{2}{l}{$\textit{  }$} & \multicolumn{1}{l}{$\textsc{CostGen}\textit{             }$} & \multicolumn{1}{l}{$\alpha$} & \multicolumn{2}{l}{$\textit{}$}\\
\multicolumn{4}{l}{$\emph{genericLessArbitraryMonoid}\textit{  }\textit{=}$} & \multicolumn{6}{l}{$\textit{
}$}\\
\multicolumn{2}{p{2ex}}{  } & \multicolumn{8}{l}{$\emph{pure}\textit{ }\emptyset\textit{ }\textit{\$\$\$?}\textit{ }\emph{genericLessArbitrary}$}\\

\end{tabular}

However we want to have fully generic
implementation that chooses the cheapest
constructor even though the datatype
does not have monoid instance.

\hypertarget{class-for-budget-conscious}{%
\subsubsection{Class for
budget-conscious}\label{class-for-budget-conscious}}

When the budget is low, we need to find
the least costly constructor each time.

So to implement it as a type class
\texttt{GLessArbitrary} that is
implemented for parts of the
\texttt{Generic} \texttt{Rep}resentation
type, we will implement two methods:

\begin{enumerate}
\def\labelenumi{\arabic{enumi}.}
\tightlist
\item
  \texttt{gLessArbitrary} is used for
  normal random data generation
\item
  \texttt{cheapest} is used when we run
  out of budget
\end{enumerate}

\setlength{\tabcolsep}{1pt}
\begin{tabular}{llllllllllllllllllllll}
\multicolumn{5}{l}{$\textbf{class}\textit{ }\textsc{GLessArbitrary}$} & \multicolumn{7}{l}{$\textit{ }\emph{datatype}$} & \multicolumn{8}{l}{$\textit{ }\textbf{where}\textit{
}$}\\
\multicolumn{2}{p{2ex}}{  } & \multicolumn{1}{l}{$\emph{gLessArbitrary}\textit{ }$} & \multicolumn{1}{l}{$\textit{::}$} & \multicolumn{1}{l}{$\textit{ }$} & \multicolumn{5}{l}{$\textsc{CostGen}$} & \multicolumn{1}{l}{$\textit{ }$} & \multicolumn{1}{l}{$\textit{(}$} & \multicolumn{2}{l}{$\emph{datatype}$} & \multicolumn{1}{l}{$\textit{ }$} & \multicolumn{1}{l}{$\emph{p}$} & \multicolumn{1}{l}{$\textit{)}$} & \multicolumn{3}{l}{$\textit{
}$}\\
\multicolumn{2}{p{2ex}}{  } & \multicolumn{1}{l}{$\emph{cheapest}\textit{       }$} & \multicolumn{1}{l}{$\textit{::}$} & \multicolumn{1}{l}{$\textit{ }$} & \multicolumn{5}{l}{$\textsc{CostGen}$} & \multicolumn{1}{l}{$\textit{ }$} & \multicolumn{1}{l}{$\textit{(}$} & \multicolumn{2}{l}{$\emph{datatype}$} & \multicolumn{1}{l}{$\textit{ }$} & \multicolumn{1}{l}{$\emph{p}$} & \multicolumn{1}{l}{$\textit{)}$} & \multicolumn{3}{l}{$\textit{
}$}\\
\multicolumn{20}{l}{$\textit{}$}\\
\multicolumn{8}{l}{$\emph{genericLessArbitrary}\textit{ }\textit{::}\textit{ }$} & \multicolumn{1}{l}{$\textit{(}$} & \multicolumn{9}{l}{$\textsc{Generic}\textit{             }$} & \multicolumn{1}{l}{$\alpha$} & \multicolumn{1}{l}{$\textit{
}$}\\
\multicolumn{8}{p{24ex}}{                        } & \multicolumn{1}{l}{$\textit{,}$} & \multicolumn{9}{l}{$\textsc{GLessArbitrary}\textit{ }\textit{(}\textsc{Rep}\textit{ }$} & \multicolumn{1}{l}{$\alpha$} & \multicolumn{1}{l}{$\textit{)}\textit{)}\textit{
}$}\\
\multicolumn{6}{p{21ex}}{                     } & \multicolumn{1}{l}{$\textit{=>}$} & \multicolumn{2}{l}{$\textit{  }$} & \multicolumn{4}{l}{$\textsc{CostGen}$} & \multicolumn{5}{l}{$\textit{             }$} & \multicolumn{1}{l}{$\alpha$} & \multicolumn{1}{l}{$\textit{}$}\\
\end{tabular}

\hypertarget{helpful-type-family}{%
\subsubsection{Helpful type
family}\label{helpful-type-family}}

First we need to compute minimum cost of
the in each branch of the type
representation. Instead of calling it
\emph{minimum cost}, we call this
function \texttt{Cheapness}.

For this we need to implement minimum
function at the type level:

\setlength{\tabcolsep}{1pt}
\begin{tabular}{llllllllllllllllllllllll}
\multicolumn{3}{l}{$\textbf{type}\textit{ }$} & \multicolumn{1}{l}{$\textbf{family}$} & \multicolumn{1}{l}{$\textit{ }$} & \multicolumn{10}{l}{$\textsc{Min}\textit{ }\emph{m}\textit{ }\emph{n}\textit{ }\textbf{where}$} & \multicolumn{7}{l}{$\textit{
}$}\\
\multicolumn{2}{p{2ex}}{  } & \multicolumn{1}{l}{$\textsc{Min}$} & \multicolumn{1}{l}{$\textit{ }\emph{m}\textit{ }\emph{n}\textit{ }\textit{=}$} & \multicolumn{1}{l}{$\textit{ }$} & \multicolumn{10}{l}{$\textsc{ChooseSmaller}$} & \multicolumn{7}{l}{$\textit{ }\textit{(}\textsc{CmpNat}\textit{ }\emph{m}\textit{ }\emph{n}\textit{)}\textit{ }\emph{m}\textit{ }\emph{n}\textit{
}$}\\
\multicolumn{22}{l}{$\textit{}$}\\
\multicolumn{16}{l}{$\textbf{type}\textit{ }\textbf{family}\textit{ }\textsc{ChooseSmaller}\textit{ }$} & \multicolumn{1}{l}{$\textit{(}$} & \multicolumn{1}{l}{$\emph{o}$} & \multicolumn{1}{l}{$\textit{::}$} & \multicolumn{3}{l}{$\textsc{Ordering}\textit{)}\textit{
}$}\\
\multicolumn{16}{p{26ex}}{                          } & \multicolumn{1}{l}{$\textit{(}$} & \multicolumn{1}{l}{$\emph{m}$} & \multicolumn{1}{l}{$\textit{::}$} & \multicolumn{1}{l}{$\textsc{Nat}$} & \multicolumn{1}{l}{$\textit{)}$} & \multicolumn{1}{l}{$\textit{
}$}\\
\multicolumn{16}{p{26ex}}{                          } & \multicolumn{1}{l}{$\textit{(}$} & \multicolumn{1}{l}{$\emph{n}$} & \multicolumn{1}{l}{$\textit{::}$} & \multicolumn{1}{l}{$\textsc{Nat}$} & \multicolumn{1}{l}{$\textit{)}$} & \multicolumn{1}{l}{$\textit{ }\textbf{where}\textit{ 
}$}\\
\end{tabular}

so we can choose the cheapest\^{}{[}We
could add instances for :

\setlength{\tabcolsep}{1pt}
\begin{tabular}{llllllllllllllllllllll}
\multicolumn{5}{l}{$\textbf{type}\textit{ }\textbf{family}$} & \multicolumn{1}{l}{$\textit{ }$} & \multicolumn{8}{l}{$\textsc{Cheapness}$} & \multicolumn{1}{l}{$\textit{ }\alpha$} & \multicolumn{1}{l}{$\textit{ }$} & \multicolumn{4}{l}{$\textit{::}\textit{ }\textsc{Nat}\textit{ }\textbf{where}\textit{
}$}\\
\multicolumn{2}{p{2ex}}{  } & \multicolumn{3}{l}{$\textsc{Cheapness}$} & \multicolumn{1}{l}{$\textit{ }$} & \multicolumn{5}{l}{$\textit{(}\alpha\textit{ }\textit{:*:}$} & \multicolumn{1}{l}{$\textit{ }$} & \multicolumn{1}{l}{$\beta$} & \multicolumn{1}{l}{$\textit{)}$} & \multicolumn{1}{l}{$\textit{  }$} & \multicolumn{1}{l}{$\textit{=}$} & \multicolumn{4}{l}{$\textit{
}$}\\
\multicolumn{4}{p{9ex}}{         } & \multicolumn{7}{l}{$\textsc{Cheapness}$} & \multicolumn{1}{l}{$\textit{ }$} & \multicolumn{1}{l}{$\alpha$} & \multicolumn{1}{l}{$\textit{ }$} & \multicolumn{1}{l}{$\textit{+}\textit{ }$} & \multicolumn{5}{l}{$\textsc{Cheapness}\textit{ }\beta\textit{
}$}\\
\multicolumn{2}{p{2ex}}{  } & \multicolumn{9}{l}{$\textsc{Cheapness}\textit{ }\textit{(}\alpha\textit{ }\textit{:+:}$} & \multicolumn{1}{l}{$\textit{ }$} & \multicolumn{1}{l}{$\beta$} & \multicolumn{1}{l}{$\textit{)}$} & \multicolumn{1}{l}{$\textit{  }$} & \multicolumn{5}{l}{$\textit{=}\textit{
}$}\\
\multicolumn{3}{p{4ex}}{    } & \multicolumn{8}{l}{$\textsc{Min}\textit{ }\textit{(}\textsc{Cheapness}$} & \multicolumn{1}{l}{$\textit{ }$} & \multicolumn{1}{l}{$\alpha$} & \multicolumn{1}{l}{$\textit{)}$} & \multicolumn{1}{l}{$\textit{ }\textit{(}$} & \multicolumn{5}{l}{$\textsc{Cheapness}\textit{ }\beta\textit{)}\textit{
}$}\\
\multicolumn{2}{p{2ex}}{  } & \multicolumn{18}{l}{$\textsc{Cheapness}\textit{  }\textsc{U1}\textit{                      }\textit{=}\textit{ }0\textit{
}$}\\
\multicolumn{2}{p{2ex}}{  } & \multicolumn{7}{l}{$\ll\emph{flat}\textit{-}\emph{types}\mathbin{>\!\!\!>}$} & \multicolumn{11}{l}{$\textit{
}$}\\
\end{tabular}

Since we are only interested in
recursive types that can potentially
blow out our budget, we can also add
cases for flat types since they seem the
cheapest:

\setlength{\tabcolsep}{1pt}
\begin{tabular}{lllllllllllllllllllll}
\multicolumn{1}{l}{$\textsc{Cheapness}$} & \multicolumn{1}{l}{$\textit{ }$} & \multicolumn{1}{l}{$\textit{(}$} & \multicolumn{1}{l}{$\textsc{S1}$} & \multicolumn{1}{l}{$\textit{ }$} & \multicolumn{1}{l}{$\alpha$} & \multicolumn{1}{l}{$\textit{ }$} & \multicolumn{1}{l}{$\textit{(}$} & \multicolumn{1}{l}{$\textsc{Rec0}$} & \multicolumn{1}{l}{$\textit{ }$} & \multicolumn{1}{l}{$\textsc{Int}\textit{       }$} & \multicolumn{1}{l}{$\textit{)}$} & \multicolumn{1}{l}{$\textit{)}$} & \multicolumn{1}{l}{$\textit{ }$} & \multicolumn{1}{l}{$\textit{=}$} & \multicolumn{1}{l}{$\textit{ }$} & \multicolumn{1}{l}{$0$} & \multicolumn{1}{l}{$\textit{}$}\\
\multicolumn{1}{l}{$\textsc{Cheapness}$} & \multicolumn{1}{l}{$\textit{ }$} & \multicolumn{1}{l}{$\textit{(}$} & \multicolumn{1}{l}{$\textsc{S1}$} & \multicolumn{1}{l}{$\textit{ }$} & \multicolumn{1}{l}{$\alpha$} & \multicolumn{1}{l}{$\textit{ }$} & \multicolumn{1}{l}{$\textit{(}$} & \multicolumn{1}{l}{$\textsc{Rec0}$} & \multicolumn{1}{l}{$\textit{ }$} & \multicolumn{1}{l}{$\textsc{Scientific}$} & \multicolumn{1}{l}{$\textit{)}$} & \multicolumn{1}{l}{$\textit{)}$} & \multicolumn{1}{l}{$\textit{ }$} & \multicolumn{1}{l}{$\textit{=}$} & \multicolumn{1}{l}{$\textit{ }$} & \multicolumn{1}{l}{$0$} & \multicolumn{1}{l}{$\textit{}$}\\
\multicolumn{1}{l}{$\textsc{Cheapness}$} & \multicolumn{1}{l}{$\textit{ }$} & \multicolumn{1}{l}{$\textit{(}$} & \multicolumn{1}{l}{$\textsc{S1}$} & \multicolumn{1}{l}{$\textit{ }$} & \multicolumn{1}{l}{$\alpha$} & \multicolumn{1}{l}{$\textit{ }$} & \multicolumn{1}{l}{$\textit{(}$} & \multicolumn{1}{l}{$\textsc{Rec0}$} & \multicolumn{1}{l}{$\textit{ }$} & \multicolumn{1}{l}{$\textsc{Double}\textit{    }$} & \multicolumn{1}{l}{$\textit{)}$} & \multicolumn{1}{l}{$\textit{)}$} & \multicolumn{1}{l}{$\textit{ }$} & \multicolumn{1}{l}{$\textit{=}$} & \multicolumn{1}{l}{$\textit{ }$} & \multicolumn{1}{l}{$0$} & \multicolumn{1}{l}{$\textit{}$}\\
\end{tabular}

\hypertarget{base-case-for-each-datatype}{%
\subsubsection{Base case for each
datatype}\label{base-case-for-each-datatype}}

For each datatype, we first write a
skeleton code that first spends a coin,
and then checks whether we have enough
funds to go on expensive path, or we are
beyond our allocation and need to
generate from among the cheapest
possible options.

\setlength{\tabcolsep}{1pt}
\begin{tabular}{llllllllllllllll}
\multicolumn{5}{l}{$\textbf{instance}$} & \multicolumn{1}{l}{$\textit{ }$} & \multicolumn{5}{l}{$\textsc{GLessArbitrary}$} & \multicolumn{1}{l}{$\textit{       }$} & \multicolumn{1}{l}{$\emph{f}$} & \multicolumn{1}{l}{$\textit{
}$}\\
\multicolumn{4}{p{6ex}}{      } & \multicolumn{1}{l}{$\textit{=>}$} & \multicolumn{1}{l}{$\textit{ }$} & \multicolumn{5}{l}{$\textsc{GLessArbitrary}$} & \multicolumn{1}{l}{$\textit{ }\textit{(}\textsc{D1}\textit{ }\emph{m}\textit{ }$} & \multicolumn{1}{l}{$\emph{f}$} & \multicolumn{1}{l}{$\textit{)}\textit{ }\textbf{where}\textit{ 
}$}\\
\multicolumn{2}{p{2ex}}{  } & \multicolumn{12}{l}{$\emph{gLessArbitrary}\textit{ }\textit{=}\textit{ }\textbf{do}\textit{
}$}\\
\multicolumn{3}{p{4ex}}{    } & \multicolumn{4}{l}{$\emph{spend}\textit{ }$} & \multicolumn{1}{l}{$1$} & \multicolumn{6}{l}{$\textit{
}$}\\
\multicolumn{3}{p{4ex}}{    } & \multicolumn{4}{l}{$\textsc{M1}\textit{ }\mathbin{\ooalign{\raise.29ex\hbox{$\scriptscriptstyle\$$}\cr\hss$\!\lozenge$\hss}}$} & \multicolumn{1}{l}{$\textit{ }$} & \multicolumn{1}{l}{$\textit{(}$} & \multicolumn{1}{l}{$\emph{cheapest}$} & \multicolumn{4}{l}{$\textit{ }\textit{\$\$\$?}\textit{ }\emph{gLessArbitrary}\textit{)}\textit{
}$}\\
\multicolumn{2}{p{2ex}}{  } & \multicolumn{5}{l}{$\emph{cheapest}$} & \multicolumn{1}{l}{$\textit{ }$} & \multicolumn{1}{l}{$\textit{=}$} & \multicolumn{1}{l}{$\textit{ }\textsc{M1}\textit{ }\mathbin{\ooalign{\raise.29ex\hbox{$\scriptscriptstyle\$$}\cr\hss$\!\lozenge$\hss}}\textit{ }$} & \multicolumn{4}{l}{$\emph{cheapest}$}\\

\end{tabular}

\hypertarget{skipping-over-other-metadata}{%
\subsubsection{Skipping over other
metadata}\label{skipping-over-other-metadata}}

First we safely ignore metadata by
writing an instance:

\setlength{\tabcolsep}{1pt}
\begin{tabular}{lllllllllllllllll}
\multicolumn{4}{l}{$\textbf{instance}$} & \multicolumn{1}{l}{$\textit{ }$} & \multicolumn{4}{l}{$\textsc{GLessArbitrary}$} & \multicolumn{4}{l}{$\textit{           }$} & \multicolumn{1}{l}{$\emph{f}$} & \multicolumn{1}{l}{$\textit{
}$}\\
\multicolumn{3}{p{6ex}}{      } & \multicolumn{1}{l}{$\textit{=>}$} & \multicolumn{1}{l}{$\textit{ }$} & \multicolumn{4}{l}{$\textsc{GLessArbitrary}$} & \multicolumn{1}{l}{$\textit{ }$} & \multicolumn{3}{l}{$\textit{(}\textsc{G.C1}\textit{ }\gamma\textit{ }\emph{f}\textit{)}$} & \multicolumn{1}{l}{$\textit{ }$} & \multicolumn{1}{l}{$\textbf{where}\textit{
}$}\\
\multicolumn{2}{p{2ex}}{  } & \multicolumn{4}{l}{$\emph{gLessArbitrary}\textit{ }$} & \multicolumn{1}{l}{$\textit{=}$} & \multicolumn{1}{l}{$\textit{ }$} & \multicolumn{1}{l}{$\textsc{G.M1}$} & \multicolumn{1}{l}{$\textit{ }$} & \multicolumn{1}{l}{$\mathbin{\ooalign{\raise.29ex\hbox{$\scriptscriptstyle\$$}\cr\hss$\!\lozenge$\hss}}$} & \multicolumn{1}{l}{$\textit{ }$} & \multicolumn{3}{l}{$\emph{gLessArbitrary}\textit{
}$}\\
\multicolumn{2}{p{2ex}}{  } & \multicolumn{4}{l}{$\emph{cheapest}\textit{       }$} & \multicolumn{1}{l}{$\textit{=}$} & \multicolumn{1}{l}{$\textit{ }$} & \multicolumn{1}{l}{$\textsc{G.M1}$} & \multicolumn{1}{l}{$\textit{ }$} & \multicolumn{1}{l}{$\mathbin{\ooalign{\raise.29ex\hbox{$\scriptscriptstyle\$$}\cr\hss$\!\lozenge$\hss}}$} & \multicolumn{1}{l}{$\textit{ }$} & \multicolumn{3}{l}{$\emph{cheapest}\textit{
}$}\\
\multicolumn{15}{l}{$\textit{}$}\\
\multicolumn{4}{l}{$\textbf{instance}$} & \multicolumn{1}{l}{$\textit{ }$} & \multicolumn{4}{l}{$\textsc{GLessArbitrary}$} & \multicolumn{4}{l}{$\textit{           }$} & \multicolumn{1}{l}{$\emph{f}$} & \multicolumn{1}{l}{$\textit{
}$}\\
\end{tabular}

\hypertarget{counting-constructors}{%
\subsubsection{Counting
constructors}\label{counting-constructors}}

In order to give equal draw chance for
each constructor, we need to count
number of constructors in each branch of
sum type \texttt{:+:} so we can generate
each constructor with the same
frequency:

\setlength{\tabcolsep}{1pt}
\begin{tabular}{llllllllllllllll}
\multicolumn{5}{l}{$\textbf{type}\textit{ }\textbf{family}$} & \multicolumn{1}{l}{$\textit{ }$} & \multicolumn{1}{l}{$\textsc{SumLen}$} & \multicolumn{1}{l}{$\textit{ }$} & \multicolumn{1}{l}{$\alpha$} & \multicolumn{1}{l}{$\textit{ }$} & \multicolumn{2}{l}{$\textit{::}$} & \multicolumn{1}{l}{$\textit{ }\textsc{Nat}\textit{ }\textbf{where}$} & \multicolumn{1}{l}{$\textit{
}$}\\
\multicolumn{2}{p{2ex}}{  } & \multicolumn{1}{l}{$\textsc{SumLen}$} & \multicolumn{1}{l}{$\textit{ }\textit{(}$} & \multicolumn{1}{l}{$\alpha$} & \multicolumn{1}{l}{$\textit{ }$} & \multicolumn{1}{l}{$\textit{G.:+:}\textit{ }$} & \multicolumn{1}{l}{$\beta$} & \multicolumn{1}{l}{$\textit{)}$} & \multicolumn{1}{l}{$\textit{ }$} & \multicolumn{1}{l}{$\textit{=}$} & \multicolumn{1}{l}{$\textit{ }$} & \multicolumn{1}{l}{$\textsc{SumLen}\textit{ }\alpha\textit{ }\textit{+}$} & \multicolumn{1}{l}{$\textit{ }\textsc{SumLen}\textit{ }\beta\textit{
}$}\\
\multicolumn{2}{p{2ex}}{  } & \multicolumn{1}{l}{$\textsc{SumLen}$} & \multicolumn{1}{l}{$\textit{  }$} & \multicolumn{1}{l}{$\alpha$} & \multicolumn{5}{l}{$\textit{          }$} & \multicolumn{1}{l}{$\textit{=}$} & \multicolumn{1}{l}{$\textit{ }$} & \multicolumn{2}{l}{$1$}\\

\end{tabular}

\hypertarget{base-cases-for-glessarbitrary}{%
\subsubsection{\texorpdfstring{Base
cases for
\texttt{GLessArbitrary}}{Base cases for GLessArbitrary}}\label{base-cases-for-glessarbitrary}}

Now we are ready to define the instances
of \texttt{GLessArbitrary} class.

We start with base cases
\texttt{GLessArbitrary} for types with
the same representation as unit type has
only one result:

\setlength{\tabcolsep}{1pt}
\begin{tabular}{lllllllllll}
\multicolumn{6}{l}{$\textbf{instance}\textit{ }\textsc{GLessArbitrary}$} & \multicolumn{1}{l}{$\textit{ }$} & \multicolumn{1}{l}{$\textsc{G.U1}$} & \multicolumn{1}{l}{$\textit{ }\textbf{where}\textit{
}$}\\
\multicolumn{2}{p{2ex}}{  } & \multicolumn{1}{l}{$\emph{gLessArbitrary}\textit{ }$} & \multicolumn{1}{l}{$\textit{=}$} & \multicolumn{1}{l}{$\textit{ }$} & \multicolumn{1}{l}{$\emph{pure}$} & \multicolumn{1}{l}{$\textit{ }$} & \multicolumn{1}{l}{$\textsc{G.U1}$} & \multicolumn{1}{l}{$\textit{
}$}\\
\multicolumn{2}{p{2ex}}{  } & \multicolumn{1}{l}{$\emph{cheapest}\textit{       }$} & \multicolumn{1}{l}{$\textit{=}$} & \multicolumn{1}{l}{$\textit{ }$} & \multicolumn{1}{l}{$\emph{pure}$} & \multicolumn{1}{l}{$\textit{ }$} & \multicolumn{2}{l}{$\textsc{G.U1}$}\\

\end{tabular}

For the product of, we descend down the
product of to reach each field, and then
assemble the result:

\setlength{\tabcolsep}{1pt}
\begin{tabular}{llllllllllllllllll}
\multicolumn{4}{l}{$\textbf{instance}\textit{ }$} & \multicolumn{1}{l}{$\textit{(}$} & \multicolumn{1}{l}{$\textsc{GLessArbitrary}$} & \multicolumn{10}{l}{$\textit{  }\alpha\textit{
}$}\\
\multicolumn{4}{p{9ex}}{         } & \multicolumn{1}{l}{$\textit{,}$} & \multicolumn{1}{l}{$\textsc{GLessArbitrary}$} & \multicolumn{6}{l}{$\textit{          }$} & \multicolumn{1}{l}{$\beta$} & \multicolumn{1}{l}{$\textit{)}$} & \multicolumn{2}{l}{$\textit{
}$}\\
\multicolumn{3}{p{6ex}}{      } & \multicolumn{2}{l}{$\textit{=>}\textit{  }$} & \multicolumn{1}{l}{$\textsc{GLessArbitrary}$} & \multicolumn{1}{l}{$\textit{ }$} & \multicolumn{1}{l}{$\textit{(}$} & \multicolumn{1}{l}{$\alpha$} & \multicolumn{3}{l}{$\textit{ }\textit{G.:*:}\textit{ }$} & \multicolumn{1}{l}{$\beta$} & \multicolumn{1}{l}{$\textit{)}$} & \multicolumn{2}{l}{$\textit{ }\textbf{where}\textit{
}$}\\
\multicolumn{2}{p{2ex}}{  } & \multicolumn{5}{l}{$\emph{gLessArbitrary}\textit{ }\textit{=}\textit{ }\textit{(}\textit{G.:*:}$} & \multicolumn{1}{l}{$\textit{)}$} & \multicolumn{1}{l}{$\textit{ }$} & \multicolumn{1}{l}{$\mathbin{\ooalign{\raise.29ex\hbox{$\scriptscriptstyle\$$}\cr\hss$\!\lozenge$\hss}}$} & \multicolumn{1}{l}{$\textit{ }$} & \multicolumn{4}{l}{$\emph{gLessArbitrary}$} & \multicolumn{1}{l}{$\textit{
}$}\\
\multicolumn{9}{p{27ex}}{                           } & \multicolumn{1}{l}{$\mathbin{\ooalign{\raise.37ex\hbox{$\scriptscriptstyle{*}$}\cr\hss$\!\lozenge$\hss}}$} & \multicolumn{1}{l}{$\textit{ }$} & \multicolumn{4}{l}{$\emph{gLessArbitrary}$} & \multicolumn{1}{l}{$\textit{
}$}\\
\multicolumn{2}{p{2ex}}{  } & \multicolumn{7}{l}{$\emph{cheapest}\textit{       }\textit{=}\textit{ }\textit{(}\textit{G.:*:}\textit{)}\textit{ }$} & \multicolumn{1}{l}{$\mathbin{\ooalign{\raise.29ex\hbox{$\scriptscriptstyle\$$}\cr\hss$\!\lozenge$\hss}}$} & \multicolumn{1}{l}{$\textit{ }$} & \multicolumn{5}{l}{$\emph{cheapest}\textit{
}$}\\
\end{tabular}

We recursively call instances of
\texttt{LessArbitrary} for the types of
fields:

\setlength{\tabcolsep}{1pt}
\begin{tabular}{llllllllllllllll}
\multicolumn{4}{l}{$\textbf{instance}$} & \multicolumn{4}{l}{$\textit{  }\textsc{LessArbitrary}$} & \multicolumn{4}{l}{$\textit{         }$} & \multicolumn{1}{l}{$\gamma$} & \multicolumn{1}{l}{$\textit{
}$}\\
\multicolumn{3}{p{6ex}}{      } & \multicolumn{1}{l}{$\textit{=>}$} & \multicolumn{4}{l}{$\textit{ }\textsc{GLessArbitrary}$} & \multicolumn{1}{l}{$\textit{ }$} & \multicolumn{3}{l}{$\textit{(}\textsc{G.K1}\textit{ }\emph{i}\textit{ }$} & \multicolumn{1}{l}{$\gamma$} & \multicolumn{1}{l}{$\textit{)}\textit{ }\textbf{where}\textit{
}$}\\
\multicolumn{2}{p{2ex}}{  } & \multicolumn{3}{l}{$\emph{gLessArbitrary}\textit{ }$} & \multicolumn{1}{l}{$\textit{=}$} & \multicolumn{1}{l}{$\textit{ }$} & \multicolumn{1}{l}{$\textsc{G.K1}$} & \multicolumn{1}{l}{$\textit{ }$} & \multicolumn{1}{l}{$\mathbin{\ooalign{\raise.29ex\hbox{$\scriptscriptstyle\$$}\cr\hss$\!\lozenge$\hss}}$} & \multicolumn{1}{l}{$\textit{ }$} & \multicolumn{3}{l}{$\emph{lessArbitrary}\textit{
}$}\\
\multicolumn{2}{p{2ex}}{  } & \multicolumn{3}{l}{$\emph{cheapest}\textit{       }$} & \multicolumn{1}{l}{$\textit{=}$} & \multicolumn{1}{l}{$\textit{ }$} & \multicolumn{1}{l}{$\textsc{G.K1}$} & \multicolumn{1}{l}{$\textit{ }$} & \multicolumn{1}{l}{$\mathbin{\ooalign{\raise.29ex\hbox{$\scriptscriptstyle\$$}\cr\hss$\!\lozenge$\hss}}$} & \multicolumn{1}{l}{$\textit{ }$} & \multicolumn{3}{l}{$\emph{lessArbitrary}$}\\

\end{tabular}

\hypertarget{selecting-the-constructor}{%
\subsubsection{Selecting the
constructor}\label{selecting-the-constructor}}

We use code for selecting the
constructor that is taken
after{[}\protect\hyperlink{ref-generic-arbitrary}{5}{]}.

\setlength{\tabcolsep}{1pt}
\begin{tabular}{lllllllllllllllllllllllllllll}
\multicolumn{7}{l}{$\textbf{instance}\textit{ }$} & \multicolumn{1}{l}{$\textit{(}$} & \multicolumn{14}{l}{$\textsc{GLessArbitrary}$} & \multicolumn{6}{l}{$\textit{      }\alpha\textit{
}$}\\
\multicolumn{7}{p{9ex}}{         } & \multicolumn{1}{l}{$\textit{,}$} & \multicolumn{14}{l}{$\textsc{GLessArbitrary}$} & \multicolumn{6}{l}{$\textit{                    }\beta\textit{
}$}\\
\multicolumn{7}{p{9ex}}{         } & \multicolumn{1}{l}{$\textit{,}$} & \multicolumn{7}{l}{$\textsc{KnownNat}$} & \multicolumn{1}{l}{$\textit{ }$} & \multicolumn{2}{l}{$\textit{(}$} & \multicolumn{5}{l}{$\textsc{SumLen}$} & \multicolumn{5}{l}{$\textit{    }\alpha\textit{)}\textit{
}$}\\
\multicolumn{7}{p{9ex}}{         } & \multicolumn{1}{l}{$\textit{,}$} & \multicolumn{7}{l}{$\textsc{KnownNat}$} & \multicolumn{1}{l}{$\textit{ }$} & \multicolumn{2}{l}{$\textit{(}$} & \multicolumn{5}{l}{$\textsc{SumLen}$} & \multicolumn{5}{l}{$\textit{                  }\beta\textit{)}\textit{
}$}\\
\multicolumn{7}{p{9ex}}{         } & \multicolumn{1}{l}{$\textit{,}$} & \multicolumn{7}{l}{$\textsc{KnownNat}$} & \multicolumn{1}{l}{$\textit{ }$} & \multicolumn{2}{l}{$\textit{(}$} & \multicolumn{7}{l}{$\textsc{Cheapness}$} & \multicolumn{3}{l}{$\textit{ }\alpha\textit{)}\textit{
}$}\\
\multicolumn{7}{p{9ex}}{         } & \multicolumn{1}{l}{$\textit{,}$} & \multicolumn{7}{l}{$\textsc{KnownNat}$} & \multicolumn{1}{l}{$\textit{ }$} & \multicolumn{2}{l}{$\textit{(}$} & \multicolumn{7}{l}{$\textsc{Cheapness}$} & \multicolumn{3}{l}{$\textit{               }\beta\textit{)}\textit{
}$}\\
\multicolumn{7}{p{9ex}}{         } & \multicolumn{1}{l}{$\textit{)}$} & \multicolumn{20}{l}{$\textit{
}$}\\
\multicolumn{4}{p{6ex}}{      } & \multicolumn{3}{l}{$\textit{=>}\textit{ }$} & \multicolumn{21}{l}{$\textsc{GLessArbitrary}\textit{      }\textit{(}\alpha\textit{ }\textit{Generic.:+:}\textit{ }\beta\textit{)}\textit{ }\textbf{where}\textit{
}$}\\
\end{tabular}

\hypertarget{conclusion}{%
\section{Conclusion}\label{conclusion}}

We show how to quickly define
terminating test generators using
generic programming. This method may be
transferred to other generic programming
regimes like Featherweight Go or
Featherweight Java.

We recommend because it reduces time
spent on making test generators and
improves user experience when a data
structure with no terminating
constructors is defined.

\hypertarget{bibliography}{%
\section{Bibliography}\label{bibliography}}

\hypertarget{refs}{}
\begin{cslreferences}
\leavevmode\hypertarget{ref-quickcheck}{}%
{[}1{]} Claessen, K. and Hughes, J.
2000. QuickCheck: A lightweight tool for
random testing of haskell programs.
\emph{SIGPLAN Not.} 35, 9 (Sep. 2000),
268--279.
DOI:\url{https://doi.org/10.1145/357766.351266}.

\leavevmode\hypertarget{ref-quickCheck}{}%
{[}2{]} Claessen, K. and Hughes, J.
2000. QuickCheck: A lightweight tool for
random testing of haskell programs.
\emph{ICFP '00: Proceedings of the fifth
acm sigplan international conference on
functional programming} (New York, NY,
USA, 2000), 268--279.

\leavevmode\hypertarget{ref-compilersALaCarte}{}%
{[}3{]} Day, L.E. and Hutton, G. 2013.
Compilation à la Carte.
\emph{Proceedings of the 25th Symposium
on Implementation and Application of
Functional Languages} (Nijmegen, The
Netherlands, 2013).

\leavevmode\hypertarget{ref-entangled}{}%
{[}4{]} EnTangleD: A bi-directional
literate programming tool: 2019.
\emph{\url{https://blog.esciencecenter.nl/entangled-1744448f4b9f}}.

\leavevmode\hypertarget{ref-generic-arbitrary}{}%
{[}5{]} generic-arbitrary: Generic
implementation for QuickCheck's
Arbitrary: 2017.
\emph{\url{https://hackage.haskell.org/package/generic-arbitrary-0.1.0/docs/src/Test-QuickCheck-Arbitrary-Generic.html\#genericArbitrary}}.

\leavevmode\hypertarget{ref-validity}{}%
{[}6{]} genvalidity-property: Standard
properties for functions on `Validity`
types: 2018.
\emph{\url{https://hackage.haskell.org/package/generic-arbitrary-0.1.0/docs/src/Test-QuickCheck-Arbitrary-Generic.html\#genericArbitrary}}.

\leavevmode\hypertarget{ref-composing-monads}{}%
{[}7{]} Jones, M.P. and Duponcheel, L.
1993. \emph{Composing monads}.

\leavevmode\hypertarget{ref-literate-programming}{}%
{[}8{]} Knuth, D.E. 1984. Literate
programming. \emph{Comput. J.} 27, 2
(May 1984), 97--111.
DOI:\url{https://doi.org/10.1093/comjnl/27.2.97}.

\leavevmode\hypertarget{ref-pandoc}{}%
{[}9{]} Pandoc: A universal document
converter: 2000.
\emph{\url{https://pandoc.org}}.

\leavevmode\hypertarget{ref-quickcheck-manual}{}%
{[}10{]} QuickCheck: An Automatic
Testing Tool for Haskell:
\emph{\url{http://www.cse.chalmers.se/~rjmh/QuickCheck/manual_body.html\#16}}.

\leavevmode\hypertarget{ref-haskell-stack}{}%
{[}11{]} stack 0.1 released:.
\end{cslreferences}

\hypertarget{appendix-module-headers}{%
\section*{Appendix: Module
headers}\label{appendix-module-headers}}
\addcontentsline{toc}{section}{Appendix:
Module headers}

\setlength{\tabcolsep}{1pt}
\begin{tabular}{lllllllllllllllllllllllllllllllllllllll}
\multicolumn{32}{l}{$\textit{\{-\# language DefaultSignatures     \#-\}}$} & \multicolumn{5}{l}{$\textit{}$}\\
\multicolumn{32}{l}{$\textit{\{-\# language FlexibleInstances     \#-\}}$} & \multicolumn{5}{l}{$\textit{}$}\\
\multicolumn{32}{l}{$\textit{\{-\# language FlexibleContexts      \#-\}}$} & \multicolumn{5}{l}{$\textit{}$}\\
\multicolumn{36}{l}{$\textit{\{-\# language GeneralizedNewtypeDeriving \#-\}}$} & \multicolumn{1}{l}{$\textit{}$}\\
\multicolumn{32}{l}{$\textit{\{-\# language Rank2Types            \#-\}}$} & \multicolumn{5}{l}{$\textit{}$}\\
\multicolumn{32}{l}{$\textit{\{-\# language PolyKinds             \#-\}}$} & \multicolumn{5}{l}{$\textit{}$}\\
\multicolumn{32}{l}{$\textit{\{-\# language MultiParamTypeClasses \#-\}}$} & \multicolumn{5}{l}{$\textit{}$}\\
\multicolumn{32}{l}{$\textit{\{-\# language MultiWayIf            \#-\}}$} & \multicolumn{5}{l}{$\textit{}$}\\
\end{tabular}

\hypertarget{appendix-lifting-classic-arbitrary-functions}{%
\section*{\texorpdfstring{Appendix:
lifting classic \texttt{Arbitrary}
functions}{Appendix: lifting classic Arbitrary functions}}\label{appendix-lifting-classic-arbitrary-functions}}
\addcontentsline{toc}{section}{Appendix:
lifting classic \texttt{Arbitrary}
functions}

Below are functions and instances that
are lightly adjusted variants of
original implementations in
\texttt{QuickCheck}{[}\protect\hyperlink{ref-quickcheck}{1}{]}

\setlength{\tabcolsep}{1pt}
\begin{tabular}{lllllllllllllllllllll}
\multicolumn{5}{l}{$\textbf{instance}$} & \multicolumn{1}{l}{$\textit{ }$} & \multicolumn{6}{l}{$\textsc{LessArbitrary}$} & \multicolumn{2}{l}{$\textit{  }$} & \multicolumn{1}{l}{$\alpha$} & \multicolumn{4}{l}{$\textit{
}$}\\
\multicolumn{4}{p{6ex}}{      } & \multicolumn{1}{l}{$\textit{=>}$} & \multicolumn{1}{l}{$\textit{ }$} & \multicolumn{6}{l}{$\textsc{LessArbitrary}$} & \multicolumn{1}{l}{$\textit{ }$} & \multicolumn{1}{l}{$\textit{[}$} & \multicolumn{1}{l}{$\alpha$} & \multicolumn{1}{l}{$\textit{]}$} & \multicolumn{3}{l}{$\textit{ }\textbf{where}\textit{
}$}\\
\multicolumn{2}{p{2ex}}{  } & \multicolumn{10}{l}{$\emph{lessArbitrary}\textit{ }\textit{=}\textit{ }\emph{pure}$} & \multicolumn{1}{l}{$\textit{ }$} & \multicolumn{1}{l}{$\textit{[}$} & \multicolumn{1}{l}{$\textit{]}$} & \multicolumn{1}{l}{$\textit{ }$} & \multicolumn{3}{l}{$\textit{\$\$\$?}\textit{ }\textbf{do}\textit{
}$}\\
\multicolumn{3}{p{4ex}}{    } & \multicolumn{4}{l}{$\emph{budget}\textit{ }$} & \multicolumn{12}{l}{$\textit{<-}\textit{ }\emph{currentBudget}\textit{
}$}\\
\multicolumn{3}{p{4ex}}{    } & \multicolumn{3}{l}{$\emph{len}\textit{  }$} & \multicolumn{1}{l}{$\textit{<-}$} & \multicolumn{1}{l}{$\textit{ }$} & \multicolumn{3}{l}{$\emph{choose}\textit{ }\textit{(}$} & \multicolumn{8}{l}{$1\textit{,}\emph{fromEnum}\textit{ }\emph{budget}\textit{)}\textit{
}$}\\
\multicolumn{3}{p{4ex}}{    } & \multicolumn{3}{l}{$\emph{spend}$} & \multicolumn{1}{l}{$\textit{ }\textit{\$}$} & \multicolumn{1}{l}{$\textit{ }$} & \multicolumn{2}{l}{$\textsc{Cost}\textit{ }$} & \multicolumn{1}{l}{$\emph{len}$} & \multicolumn{8}{l}{$\textit{
}$}\\
\multicolumn{3}{p{4ex}}{    } & \multicolumn{7}{l}{$\emph{replicateM}\textit{   }$} & \multicolumn{1}{l}{$\emph{len}$} & \multicolumn{8}{l}{$\textit{ }\emph{lessArbitrary}\textit{
}$}\\
\multicolumn{19}{l}{$\textit{}$}\\
\end{tabular}

Remaining functions are directly copied
from
\texttt{QuickCheck}{[}\protect\hyperlink{ref-quickCheck}{2}{]},
with only adjustment being their types
and error messages:

\setlength{\tabcolsep}{1pt}
\begin{tabular}{lllllllllllllllllllllllll}
\multicolumn{4}{l}{$\emph{forAll}$} & \multicolumn{1}{l}{$\textit{ }$} & \multicolumn{3}{l}{$\textit{::}\textit{ }$} & \multicolumn{7}{l}{$\textsc{CostGen}$} & \multicolumn{1}{l}{$\textit{ }$} & \multicolumn{2}{l}{$\alpha\textit{ }\textit{->}$} & \multicolumn{3}{l}{$\textit{ }\textit{(}\alpha$} & \multicolumn{1}{l}{$\textit{ }$} & \multicolumn{1}{l}{$\textit{->}\textit{ }\textsc{CostGen}\textit{ }\beta\textit{)}\textit{ }\textit{->}\textit{ }\textsc{CostGen}\textit{ }\beta\textit{}$}\\
\multicolumn{4}{l}{$\emph{forAll}$} & \multicolumn{1}{l}{$\textit{ }$} & \multicolumn{3}{l}{$\emph{gen}$} & \multicolumn{7}{l}{$\textit{ }\emph{prop}\textit{ }\textit{=}$} & \multicolumn{1}{l}{$\textit{ }$} & \multicolumn{2}{l}{$\emph{gen}\textit{ }$} & \multicolumn{3}{l}{$\mathbin{>\!\!\!>\!\!=}$} & \multicolumn{1}{l}{$\textit{ }$} & \multicolumn{1}{l}{$\emph{prop}\textit{
}$}\\
\multicolumn{23}{l}{$\textit{}$}\\
\multicolumn{6}{l}{$\emph{oneof}\textit{   }$} & \multicolumn{2}{l}{$\textit{::}$} & \multicolumn{1}{l}{$\textit{ }$} & \multicolumn{10}{l}{$\textsc{HasCallStack}$} & \multicolumn{4}{l}{$\textit{
}$}\\
\multicolumn{6}{p{8ex}}{        } & \multicolumn{2}{l}{$\textit{=>}$} & \multicolumn{1}{l}{$\textit{ }$} & \multicolumn{10}{l}{$\textit{[}\textsc{CostGen}\textit{ }\alpha\textit{]}\textit{ }$} & \multicolumn{4}{l}{$\textit{->}\textit{ }\textsc{CostGen}\textit{ }\alpha\textit{}$}\\
\multicolumn{6}{l}{$\emph{oneof}\textit{ }\textit{[}\textit{]}$} & \multicolumn{2}{l}{$\textit{ }\textit{=}$} & \multicolumn{1}{l}{$\textit{ }$} & \multicolumn{14}{l}{$\emph{error}\textit{
}$}\\
\multicolumn{9}{p{11ex}}{           } & \multicolumn{14}{l}{$\textit{"LessArbitrary.oneof used with empty list"}\textit{}$}\\
\end{tabular}

This key function, chooses one of the
given generators, with a weighted random
distribution. The input list must be
non-empty. Based on
QuickCheck{[}\protect\hyperlink{ref-quickcheck}{1}{]}.

\setlength{\tabcolsep}{1pt}
\begin{tabular}{lllllllllllllllllllllll}
\multicolumn{12}{l}{$\emph{frequency}\textit{   }$} & \multicolumn{1}{l}{$\textit{::}$} & \multicolumn{1}{l}{$\textit{ }$} & \multicolumn{7}{l}{$\textsc{HasCallStack}\textit{
}$}\\
\multicolumn{12}{p{12ex}}{            } & \multicolumn{1}{l}{$\textit{=>}$} & \multicolumn{1}{l}{$\textit{ }$} & \multicolumn{7}{l}{$\textit{[}\textit{(}\textsc{Int}\textit{,}\textit{ }\textsc{CostGen}\textit{ }\alpha\textit{)}\textit{]}\textit{ }\textit{->}\textit{ }\textsc{CostGen}\textit{ }\alpha\textit{}$}\\
\multicolumn{9}{l}{$\emph{frequency}$} & \multicolumn{1}{l}{$\textit{ }$} & \multicolumn{2}{l}{$\textit{[}\textit{]}$} & \multicolumn{1}{l}{$\textit{ }\textit{=}$} & \multicolumn{8}{l}{$\textit{
}$}\\
\multicolumn{3}{p{2ex}}{  } & \multicolumn{4}{l}{$\emph{error}$} & \multicolumn{2}{l}{$\textit{ }\textit{\$}$} & \multicolumn{1}{l}{$\textit{ }$} & \multicolumn{11}{l}{$\textit{"LessArbitrary.frequency "}\textit{
}$}\\
\multicolumn{7}{p{7ex}}{       } & \multicolumn{2}{l}{$\textit{++}$} & \multicolumn{1}{l}{$\textit{ }$} & \multicolumn{11}{l}{$\textit{"used with empty list"}\textit{}$}\\
\multicolumn{9}{l}{$\emph{frequency}$} & \multicolumn{1}{l}{$\textit{ }$} & \multicolumn{2}{l}{$\emph{xs}$} & \multicolumn{9}{l}{$\textit{
}$}\\
\multicolumn{3}{p{2ex}}{  } & \multicolumn{1}{l}{$\textit{|}\textit{ }$} & \multicolumn{5}{l}{$\emph{any}\textit{ }\textit{(}$} & \multicolumn{1}{l}{$\textit{<}$} & \multicolumn{1}{l}{$\textit{ }$} & \multicolumn{1}{l}{$0$} & \multicolumn{9}{l}{$\textit{)}\textit{ }\textit{(}\emph{map}\textit{ }\emph{fst}\textit{ }\emph{xs}\textit{)}\textit{ }\textit{=}\textit{
}$}\\
\multicolumn{4}{p{4ex}}{    } & \multicolumn{5}{l}{$\emph{error}$} & \multicolumn{1}{l}{$\textit{ }$} & \multicolumn{1}{l}{$\textit{\$}$} & \multicolumn{1}{l}{$\textit{ }$} & \multicolumn{9}{l}{$\textit{"LessArbitrary.frequency: "}\textit{
}$}\\
\end{tabular}

\hypertarget{appendix-test-suite}{%
\section*{Appendix: test
suite}\label{appendix-test-suite}}
\addcontentsline{toc}{section}{Appendix:
test suite}

As observed in
{[}\protect\hyperlink{ref-validity}{6}{]},
it is important to check basic
properties of \texttt{Arbitrary}
instance to guarantee that shrinking
terminates:

\setlength{\tabcolsep}{1pt}
\begin{tabular}{llllllllllllllllll}
\multicolumn{6}{l}{$\emph{shrinkCheck}\textit{ }\textit{::}$} & \multicolumn{1}{l}{$\textit{ }$} & \multicolumn{4}{l}{$\forall\textit{    }$} & \multicolumn{2}{l}{$\emph{term}$} & \multicolumn{3}{l}{$\textit{.}\textit{
}$}\\
\multicolumn{6}{p{14ex}}{              } & \multicolumn{1}{l}{$\textit{(}$} & \multicolumn{4}{l}{$\textsc{Arbitrary}\textit{ }$} & \multicolumn{2}{l}{$\emph{term}$} & \multicolumn{3}{l}{$\textit{
}$}\\
\multicolumn{6}{p{14ex}}{              } & \multicolumn{1}{l}{$\textit{,}$} & \multicolumn{4}{l}{$\textsc{Eq}\textit{        }$} & \multicolumn{2}{l}{$\emph{term}$} & \multicolumn{3}{l}{$\textit{)}\textit{
}$}\\
\multicolumn{5}{p{12ex}}{            } & \multicolumn{1}{l}{$\textit{=>}$} & \multicolumn{5}{l}{$\textit{           }$} & \multicolumn{2}{l}{$\emph{term}$} & \multicolumn{3}{l}{$\textit{
}$}\\
\multicolumn{5}{p{12ex}}{            } & \multicolumn{1}{l}{$\textit{->}$} & \multicolumn{10}{l}{$\textit{ }\textsc{Bool}\textit{}$}\\
\multicolumn{5}{l}{$\emph{shrinkCheck}\textit{ }$} & \multicolumn{3}{l}{$\emph{term}$} & \multicolumn{1}{l}{$\textit{ }$} & \multicolumn{7}{l}{$\textit{=}\textit{
}$}\\
\multicolumn{2}{p{2ex}}{  } & \multicolumn{6}{l}{$\emph{term}\textit{ }\textit{`}\emph{notElem}\textit{`}$} & \multicolumn{1}{l}{$\textit{ }$} & \multicolumn{7}{l}{$\emph{shrink}\textit{ }\emph{term}\textit{
}$}\\
\multicolumn{16}{l}{$\textit{}$}\\
\end{tabular}

For \texttt{LessArbitrary} we can also
check that empty budget results in
choosing a cheapest option, but we need
to provide a predicate that confirms
what is actually the cheapest:

\setlength{\tabcolsep}{1pt}
\begin{tabular}{llllllllllllllllllllllllll}
\multicolumn{4}{l}{$\emph{otherLaws}$} & \multicolumn{1}{l}{$\textit{ }$} & \multicolumn{2}{l}{$\textit{::}$} & \multicolumn{1}{l}{$\textit{ }$} & \multicolumn{16}{l}{$\textit{[}\textsc{Laws}\textit{]}\textit{}$}\\
\multicolumn{4}{l}{$\emph{otherLaws}$} & \multicolumn{1}{l}{$\textit{ }$} & \multicolumn{2}{l}{$\textit{=}\textit{ }$} & \multicolumn{1}{l}{$\textit{[}$} & \multicolumn{16}{l}{$\emph{lessArbitraryLaws}\textit{ }\emph{isLeaf}\textit{]}\textit{
}$}\\
\multicolumn{2}{p{2ex}}{  } & \multicolumn{22}{l}{$\textbf{where}\textit{
}$}\\
\multicolumn{3}{p{4ex}}{    } & \multicolumn{2}{l}{$\emph{isLeaf}$} & \multicolumn{1}{l}{$\textit{ }$} & \multicolumn{4}{l}{$\textit{::}\textit{ }\textsc{Tree}\textit{ }$} & \multicolumn{3}{l}{$\textsc{Int}$} & \multicolumn{1}{l}{$\textit{ }$} & \multicolumn{10}{l}{$\textit{->}\textit{ }\textsc{Bool}\textit{
}$}\\
\multicolumn{3}{p{4ex}}{    } & \multicolumn{2}{l}{$\emph{isLeaf}$} & \multicolumn{1}{l}{$\textit{ }$} & \multicolumn{1}{l}{$\textit{(}$} & \multicolumn{3}{l}{$\textsc{Leaf}\textit{   }$} & \multicolumn{1}{l}{$\textit{\_{}}$} & \multicolumn{1}{l}{$\textit{)}$} & \multicolumn{1}{l}{$\textit{ }$} & \multicolumn{1}{l}{$\textit{=}$} & \multicolumn{1}{l}{$\textit{ }$} & \multicolumn{9}{l}{$\textsc{True}\textit{
}$}\\
\multicolumn{3}{p{4ex}}{    } & \multicolumn{2}{l}{$\emph{isLeaf}$} & \multicolumn{1}{l}{$\textit{ }$} & \multicolumn{1}{l}{$\textit{(}$} & \multicolumn{3}{l}{$\textsc{Branch}\textit{ }$} & \multicolumn{1}{l}{$\textit{\_{}}$} & \multicolumn{1}{l}{$\textit{)}$} & \multicolumn{1}{l}{$\textit{ }$} & \multicolumn{1}{l}{$\textit{=}$} & \multicolumn{1}{l}{$\textit{ }$} & \multicolumn{9}{l}{$\textsc{False}\textit{
}$}\\
\multicolumn{24}{l}{$\textit{}$}\\
\end{tabular}

Again some module headers:

\setlength{\tabcolsep}{1pt}
\begin{tabular}{lllllllllll}
\multicolumn{8}{l}{$\textit{\{-\# language DataKinds             \#-\}}$} & \multicolumn{1}{l}{$\textit{}$}\\
\multicolumn{8}{l}{$\textit{\{-\# language FlexibleInstances     \#-\}}$} & \multicolumn{1}{l}{$\textit{}$}\\
\multicolumn{8}{l}{$\textit{\{-\# language Rank2Types            \#-\}}$} & \multicolumn{1}{l}{$\textit{}$}\\
\multicolumn{8}{l}{$\textit{\{-\# language MultiParamTypeClasses \#-\}}$} & \multicolumn{1}{l}{$\textit{}$}\\
\multicolumn{8}{l}{$\textit{\{-\# language ScopedTypeVariables   \#-\}}$} & \multicolumn{1}{l}{$\textit{}$}\\
\multicolumn{8}{l}{$\textit{\{-\# language TypeOperators         \#-\}}$} & \multicolumn{1}{l}{$\textit{}$}\\
\multicolumn{8}{l}{$\textit{\{-\# language UndecidableInstances  \#-\}}$} & \multicolumn{1}{l}{$\textit{}$}\\
\multicolumn{8}{l}{$\textit{\{-\# language AllowAmbiguousTypes   \#-\}}$} & \multicolumn{1}{l}{$\textit{}$}\\
\end{tabular}

And we can compare the tests with
\texttt{LessArbitrary} (which terminates
fast, linear time):

\setlength{\tabcolsep}{1pt}
\begin{tabular}{llllllllll}
\multicolumn{1}{l}{$\ll$} & \multicolumn{1}{l}{$\emph{test}$} & \multicolumn{1}{c}{$\textit{-}$} & \multicolumn{1}{l}{$\emph{file}$} & \multicolumn{1}{c}{$\textit{-}$} & \multicolumn{2}{l}{$\emph{header}\mathbin{>\!\!\!>}\textit{}$}\\
\multicolumn{1}{l}{$\ll$} & \multicolumn{1}{l}{$\emph{test}$} & \multicolumn{1}{c}{$\textit{-}$} & \multicolumn{1}{l}{$\emph{less}$} & \multicolumn{1}{c}{$\textit{-}$} & \multicolumn{2}{l}{$\emph{arbitrary}\textit{-}\emph{version}\mathbin{>\!\!\!>}\textit{
}$}\\
\multicolumn{7}{l}{$\textit{}$}\\
\multicolumn{1}{l}{$\ll$} & \multicolumn{1}{l}{$\emph{test}$} & \multicolumn{1}{c}{$\textit{-}$} & \multicolumn{3}{l}{$\emph{file}\textit{-}\emph{laws}$} & \multicolumn{1}{c}{$\mathbin{>\!\!\!>}\textit{}$}\\
\multicolumn{1}{l}{$\ll$} & \multicolumn{1}{l}{$\emph{less}$} & \multicolumn{1}{c}{$\textit{-}$} & \multicolumn{3}{l}{$\emph{arbitrary}$} & \multicolumn{1}{c}{$\textit{-}\emph{check}\mathbin{>\!\!\!>}$}\\

\end{tabular}

\hypertarget{appendix-non-terminating-test-suite}{%
\section*{Appendix: non-terminating test
suite}\label{appendix-non-terminating-test-suite}}
\addcontentsline{toc}{section}{Appendix:
non-terminating test suite}

Or with a generic \texttt{Arbitrary}
(which naturally hangs):

\setlength{\tabcolsep}{1pt}
\begin{tabular}{llllllllll}
\multicolumn{1}{l}{$\ll$} & \multicolumn{1}{l}{$\emph{test}$} & \multicolumn{1}{c}{$\textit{-}$} & \multicolumn{1}{l}{$\emph{file}$} & \multicolumn{1}{l}{$\textit{-}$} & \multicolumn{1}{l}{$\emph{header}\mathbin{>\!\!\!>}$} & \multicolumn{1}{l}{$\textit{}$}\\
\multicolumn{1}{l}{$\ll$} & \multicolumn{1}{l}{$\emph{tree}$} & \multicolumn{1}{c}{$\textit{-}$} & \multicolumn{1}{l}{$\textbf{type}$} & \multicolumn{1}{l}{$\textit{-}$} & \multicolumn{1}{l}{$\emph{typical}\textit{-}$} & \multicolumn{1}{l}{$\emph{arbitrary}\mathbin{>\!\!\!>}\textit{}$}\\
\multicolumn{4}{l}{$\emph{otherLaws}\textit{ }\textit{=}$} & \multicolumn{1}{l}{$\textit{ }$} & \multicolumn{2}{l}{$\textit{[}\textit{]}\textit{}$}\\
\multicolumn{4}{l}{$\ll\emph{test}\textit{-}\emph{file}$} & \multicolumn{1}{l}{$\textit{-}$} & \multicolumn{2}{l}{$\emph{laws}\mathbin{>\!\!\!>}$}\\

\end{tabular}

Here is the code:

\setlength{\tabcolsep}{1pt}
\begin{tabular}{llllllll}
\multicolumn{4}{l}{$\textit{\{-\# language FlexibleInstances     \#-\}}$} & \multicolumn{1}{l}{$\textit{}$}\\
\multicolumn{4}{l}{$\textit{\{-\# language Rank2Types            \#-\}}$} & \multicolumn{1}{l}{$\textit{}$}\\
\multicolumn{4}{l}{$\textit{\{-\# language MultiParamTypeClasses \#-\}}$} & \multicolumn{1}{l}{$\textit{}$}\\
\multicolumn{4}{l}{$\textit{\{-\# language ScopedTypeVariables   \#-\}}$} & \multicolumn{1}{l}{$\textit{}$}\\
\multicolumn{4}{l}{$\textit{\{-\# language TypeOperators         \#-\}}$} & \multicolumn{1}{l}{$\textit{}$}\\
\multicolumn{4}{l}{$\textit{\{-\# language UndecidableInstances  \#-\}}$} & \multicolumn{1}{l}{$\textit{}$}\\
\multicolumn{4}{l}{$\textit{\{-\# language AllowAmbiguousTypes   \#-\}}$} & \multicolumn{1}{l}{$\textit{}$}\\
\multicolumn{4}{l}{$\textit{\{-\# language DeriveGeneric         \#-\}}$} & \multicolumn{1}{l}{$\textit{}$}\\
\end{tabular}

\setlength{\tabcolsep}{1pt}
\begin{tabular}{lllllllllllll}
\multicolumn{4}{l}{$\textbf{instance}$} & \multicolumn{1}{l}{$\textit{ }$} & \multicolumn{1}{l}{$\textsc{LessArbitrary}$} & \multicolumn{3}{l}{$\textit{       }$} & \multicolumn{1}{l}{$\alpha$} & \multicolumn{1}{l}{$\textit{
}$}\\
\multicolumn{3}{p{6ex}}{      } & \multicolumn{1}{l}{$\textit{=>}$} & \multicolumn{1}{l}{$\textit{ }$} & \multicolumn{1}{l}{$\textsc{LessArbitrary}$} & \multicolumn{3}{l}{$\textit{ }\textit{(}\textsc{Tree}\textit{ }$} & \multicolumn{1}{l}{$\alpha$} & \multicolumn{1}{l}{$\textit{)}\textit{ }\textbf{where}\textit{
}$}\\
\multicolumn{11}{l}{$\textit{}$}\\
\multicolumn{4}{l}{$\textbf{instance}$} & \multicolumn{1}{l}{$\textit{ }$} & \multicolumn{2}{l}{$\textsc{LessArbitrary}\textit{   }$} & \multicolumn{1}{l}{$\alpha$} & \multicolumn{3}{l}{$\textit{
}$}\\
\multicolumn{3}{p{6ex}}{      } & \multicolumn{1}{l}{$\textit{=>}$} & \multicolumn{1}{l}{$\textit{ }$} & \multicolumn{2}{l}{$\textsc{Arbitrary}\textit{ }\textit{(}\textsc{Tree}\textit{ }$} & \multicolumn{1}{l}{$\alpha$} & \multicolumn{3}{l}{$\textit{)}\textit{ }\textbf{where}\textit{
}$}\\
\multicolumn{2}{p{2ex}}{  } & \multicolumn{9}{l}{$\emph{arbitrary}\textit{ }\textit{=}\textit{ }\emph{fasterArbitrary}$}\\

\end{tabular}

\setlength{\tabcolsep}{1pt}
\begin{tabular}{lllllllllllllllllllllll}
\multicolumn{21}{l}{$\textit{}$}\\
\multicolumn{3}{l}{$\emph{main}$} & \multicolumn{1}{l}{$\textit{ }$} & \multicolumn{2}{l}{$\textit{::}$} & \multicolumn{15}{l}{$\textit{ }\textsc{IO}\textit{ }\textit{(}\textit{)}\textit{}$}\\
\multicolumn{3}{l}{$\emph{main}$} & \multicolumn{1}{l}{$\textit{ }$} & \multicolumn{2}{l}{$\textit{=}\textit{ }$} & \multicolumn{15}{l}{$\textbf{do}\textit{
}$}\\
\multicolumn{2}{p{2ex}}{  } & \multicolumn{19}{l}{$\emph{lawsCheckMany}\textit{
}$}\\
\multicolumn{3}{p{4ex}}{    } & \multicolumn{2}{l}{$\textit{[}\textit{(}$} & \multicolumn{16}{l}{$\textit{"Tree"}\textit{,}\textit{
}$}\\
\multicolumn{5}{p{6ex}}{      } & \multicolumn{1}{l}{$\textit{[}$} & \multicolumn{1}{l}{$\emph{arbitraryLaws}\textit{ }$} & \multicolumn{1}{l}{$\textit{(}$} & \multicolumn{1}{l}{$\textsc{Proxy}$} & \multicolumn{1}{l}{$\textit{ }$} & \multicolumn{1}{l}{$\textit{::}$} & \multicolumn{1}{l}{$\textit{ }$} & \multicolumn{1}{l}{$\textsc{Proxy}$} & \multicolumn{1}{l}{$\textit{ }$} & \multicolumn{1}{l}{$\textit{(}$} & \multicolumn{1}{l}{$\textsc{Tree}$} & \multicolumn{1}{l}{$\textit{ }$} & \multicolumn{1}{l}{$\textsc{Int}$} & \multicolumn{1}{l}{$\textit{)}$} & \multicolumn{1}{l}{$\textit{)}$} & \multicolumn{1}{l}{$\textit{
}$}\\
\multicolumn{5}{p{6ex}}{      } & \multicolumn{1}{l}{$\textit{,}$} & \multicolumn{1}{l}{$\emph{eqLaws}\textit{        }$} & \multicolumn{1}{l}{$\textit{(}$} & \multicolumn{1}{l}{$\textsc{Proxy}$} & \multicolumn{1}{l}{$\textit{ }$} & \multicolumn{1}{l}{$\textit{::}$} & \multicolumn{1}{l}{$\textit{ }$} & \multicolumn{1}{l}{$\textsc{Proxy}$} & \multicolumn{1}{l}{$\textit{ }$} & \multicolumn{1}{l}{$\textit{(}$} & \multicolumn{1}{l}{$\textsc{Tree}$} & \multicolumn{1}{l}{$\textit{ }$} & \multicolumn{1}{l}{$\textsc{Int}$} & \multicolumn{1}{l}{$\textit{)}$} & \multicolumn{1}{l}{$\textit{)}$} & \multicolumn{1}{l}{$\textit{
}$}\\
\end{tabular}

\setlength{\tabcolsep}{1pt}
\begin{tabular}{llll}
\multicolumn{1}{l}{$\textit{\{-\# LANGUAGE GeneralizedNewtypeDeriving \#-\}}\textit{}$}\\
\multicolumn{1}{l}{$\textbf{module}\textit{ }\textsc{Test.LessArbitrary.Cost}\textit{ }\textbf{where}\textit{
}$}\\
\multicolumn{1}{l}{$\textit{}$}\\
\multicolumn{1}{l}{$\ll\emph{cost}\mathbin{>\!\!\!>}$}\\

\end{tabular}

\hypertarget{appendix-convenience-functions-provided-with-the-module}{%
\section*{Appendix: convenience
functions provided with the
module}\label{appendix-convenience-functions-provided-with-the-module}}
\addcontentsline{toc}{section}{Appendix:
convenience functions provided with the
module}

Then we limit our choices when budget is
tight:

\setlength{\tabcolsep}{1pt}
\begin{tabular}{llllllll}
\multicolumn{1}{l}{$\emph{currentBudget}$} & \multicolumn{1}{l}{$\textit{ }$} & \multicolumn{1}{l}{$\textit{::}$} & \multicolumn{1}{l}{$\textit{ }\textsc{CostGen}$} & \multicolumn{1}{l}{$\textit{ }\textsc{Cost}\textit{}$}\\
\multicolumn{1}{l}{$\emph{currentBudget}$} & \multicolumn{1}{l}{$\textit{ }$} & \multicolumn{1}{l}{$\textit{=}\textit{ }$} & \multicolumn{1}{l}{$\textsc{CostGen}\textit{ }$} & \multicolumn{1}{l}{$\emph{State.get}$}\\

\end{tabular}

\setlength{\tabcolsep}{1pt}
\begin{tabular}{lllllllllllllll}
\multicolumn{10}{l}{$\textit{-- unused: loop breaker message type name}$} & \multicolumn{3}{l}{$\textit{}$}\\
\multicolumn{5}{l}{$\textbf{type}\textit{ }\textbf{family}$} & \multicolumn{1}{l}{$\textit{ }$} & \multicolumn{7}{l}{$\textsc{ShowType}\textit{ }\emph{k}\textit{ }\textbf{where}\textit{
}$}\\
\multicolumn{2}{p{2ex}}{  } & \multicolumn{2}{l}{$\textsc{ShowType}$} & \multicolumn{1}{l}{$\textit{ }$} & \multicolumn{1}{l}{$\textit{(}$} & \multicolumn{2}{l}{$\textsc{D1}\textit{ }\textit{(}\textit{'}$} & \multicolumn{2}{l}{$\textsc{MetaData}\textit{ }\emph{name}\textit{ }\textit{\_{}}\textit{ }\textit{\_{}}\textit{ }\textit{\_{}}\textit{)}\textit{ }\textit{\_{}}\textit{)}\textit{ }$} & \multicolumn{1}{l}{$\textit{=}$} & \multicolumn{1}{l}{$\textit{ }$} & \multicolumn{1}{l}{$\emph{name}\textit{
}$}\\
\multicolumn{2}{p{2ex}}{  } & \multicolumn{2}{l}{$\textsc{ShowType}$} & \multicolumn{2}{l}{$\textit{  }$} & \multicolumn{2}{l}{$\emph{other}$} & \multicolumn{2}{l}{$\textit{                        }$} & \multicolumn{1}{l}{$\textit{=}$} & \multicolumn{1}{l}{$\textit{ }$} & \multicolumn{1}{l}{$\textit{"unknown type"}\textit{
}$}\\
\multicolumn{13}{l}{$\textit{}$}\\
\multicolumn{6}{l}{$\emph{showType}\textit{ }\textit{::}\textit{ }$} & \multicolumn{3}{l}{$\forall\textit{                      }$} & \multicolumn{1}{l}{$\alpha$} & \multicolumn{3}{l}{$\textit{.}\textit{
}$}\\
\multicolumn{6}{p{12ex}}{            } & \multicolumn{1}{l}{$\textit{(}$} & \multicolumn{2}{l}{$\textsc{Generic}\textit{                    }$} & \multicolumn{1}{l}{$\alpha$} & \multicolumn{3}{l}{$\textit{
}$}\\
\multicolumn{6}{p{12ex}}{            } & \multicolumn{1}{l}{$\textit{,}$} & \multicolumn{2}{l}{$\textsc{KnownSymbol}\textit{ }\textit{(}\textsc{ShowType}\textit{ }\textit{(}\textsc{Rep}\textit{ }$} & \multicolumn{1}{l}{$\alpha$} & \multicolumn{3}{l}{$\textit{)}\textit{)}\textit{)}\textit{
}$}\\
\end{tabular}

\begin{acks}                            
The text was prepared with great help of
bidirectional literate
programming{[}\protect\hyperlink{ref-literate-programming}{8}{]}
tool{[}\protect\hyperlink{ref-entangled}{4}{]},
Pandoc{[}\protect\hyperlink{ref-pandoc}{9}{]}
markdown publishing system and live
feedback from Stack file watching
option{[}\protect\hyperlink{ref-haskell-stack}{11}{]}.
\end{acks}

\bibliography{less-arbitrary.bib}

\end{document}